\documentclass[12pt]{iopart}
\usepackage{iopams}  
\usepackage{epsfig}

\newcommand{\bea}{\begin{eqnarray}}
\newcommand{\eea}{\end{eqnarray}}

\newcommand{\vect}[1]{\mathbf{#1}}

\begin{document}

\title{ Integral equations for simple fluids in a general reference 
functional approach}

\author{M. Oettel\dag  
\footnote[3]{
oettel@mf.mpg.de}
}

\address{\dag\ 
Max--Planck--Institut f\"ur Metallforschung, Heisenbergstr. 3, 70569 Stuttgart \\
Institut f\"ur Theoretische und Angewandte Physik, Universit\"at Stuttgart,
 Pfaffenwaldring 57, 70569 Stuttgart}

\begin{abstract}
 The integral equations for the correlation functions of an inhomogeneous fluid
 mixture are derived using a  functional Taylor expansion of the free energy around an
 inhomogeneous equilibrium distribution. The system of equations is closed by
 the introduction of a reference functional for the correlations beyond second order
 in the density difference from the equilibrium distribution.
 Explicit expressions are obtained  for energies required to insert 
 particles of the fluid mixture into the inhomogeneous system. 
 The approach is illustrated by the determination of
 the equation of state of a simple, truncated Lennard--Jones fluid 
 and  the analysis of the behavior of this fluid near a
 hard wall. The wall--fluid integral equation exhibits complete drying and
 the corresponding coexisting densities are in good agreement with
 those  obtained from the standard (Maxwell) construction applied to the bulk fluid. 
 Self--consistency of the approach is examined by analyzing the virial/compressibility
 routes to the equation of state and the Gibbs--Duhem relation for the bulk fluid, and the
 contact density sum rule and the Gibbs adsorption equation for the hard wall problem.
 For the bulk fluid, we find good self--consistency for stable states outside
 the critical region. For the hard wall problem, the Gibbs adsorption equation
 is fulfilled very well near phase coexistence where the adsorption is large.
 For the contact density sum rule, we find deviations of up to 20\% in the ratio
 of the contact densities predicted by the present method and predicted by the
 sum rule. These deviations are largely due to a slight disagreement between the coexisting
 density for the gas phase obtained from the Maxwell construction and from complete
 drying at the hard wall.   
\end{abstract}


\submitto{}

\maketitle

\section{Introduction}

Integral equations have become a popular and increasingly accurate tool 
in describing the thermodynamics and the bulk structure of one--component
simple liquids and, to a smaller degree of accuracy, of fluid mixtures.
In  the early, formal work (see references \cite{Mor60,Ste64,Per64} and references therein) 
the structural
ingredients (pair correlation function, direct correlation function and bridge
function) have been defined in terms of graphical expansions and two exact equations
relating these three functions have been established. The unknown third
equation which is necessary for a closed system of equations is usually referred
to as the closure relation. A number of approximations have been developed
in the past for the closure relation. Some of the more succesful ones 
for the structure and thermodynamics of a one--component liquid
are the optimized random phase approximation (ORPA) \cite{Pas98},
the hierarchical reference theory (HRT) \cite{Par95}, the self--consistent
Ornstein--Zernike approximation (SCOZA) \cite{Pin02}, the 
Martynov--Sarkisov closure \cite{Vom96} and the reference hypernetted
chain equations (RHNC) \cite{Lad83}. For a general introduction into the subject treating
some of these and other common closures, see reference \cite{Han90}.
The extension of integral equation theories to a general inhomogeneous system 
defined by an external potential
(a wall, say) is possible by incorporating the source of inhomogeneity as
a second  component in the dilute limit (the wall particle) 
and treating the originally inhomogeneous problem as a bulk (homogeneous) problem in
the two--component mixture. This generalization fails for
all of the above closures whenever the phenomenon of wetting in the
presence of the external (wall) potential is involved.  

On the one hand this failure is linked to the difficulties of defining
the free energy and chemical potential accurately and consistently 
in the two--component (wall--fluid) 
mixture. Or, whenever a free energy approximation underlying a certain closure relation
can be formulated, its form does not allow for wetting.
For example,  it can be proved that in the hard wall case 
the HNC and RHNC closures miss the phenomenon of complete drying for the above 
reason (for HNC, the proof is given in reference \cite{Eva83} and 
it can be easily extended to RHNC).
This problem was tackled tentatively in references \cite{Zho90a,Zho90b} in what is termed 
the hydrostatic HNC approach and this will be commented on in section \ref{sec:hhnc}. 

On the other hand, one is inclined to link this failure to the difficulty of incorporating
the wall as a second component. As the formal framework embodied
in the graphical expansions holds equally well for a general inhomogeneous situation,
one may abandon the mixture idea and study the one--component fluid using
the common closure relations but now for the inhomogeneous correlation
functions. These
are defined in the presence of an
external potential which necessarily entails a loss of symmetry for the correlation functions 
and thus results in  a noticeable increase
in computational power required  for a numerical solution. Nevertheless,
the problem of defining the chemical potential in the inhomogeneous situation 
still persists; for a review of possible solutions for a range of commonly
used closures  see reference \cite{Kje89}. Away from phase transitions, inhomogeneous
integral equation closures may give very accurate density profiles; for the example
of a fluid confined in a slit see reference \cite{Kje90}. However, systematic
studies of wetting and drying phenomena within this approach
have remained fragmentary, see references \cite{Nie81,McG86,Pli86,Bru87}.
Very recently and quite distinct from the usual closures, an implementation of 
HRT for the inhomogeneous problem has been presented in reference \cite{Orl04}.

Parallel to the development of integral equation approaches, density functional
models have become established as a widespread tool for the investigation of 
inhomogeneous fluids and for the study of wetting phenomena, in particular.
The first study of the latter using a `modern' density functional theory
has been reported in reference \cite{Ebn77};  for a summary of the initial development 
of the field see reference \cite{Eva92}.
In contrast to integral equations, a very simple
mean--field type of functional (with a local and therefore actually  quite inadequate
treatment of the sharply repulsive interatomic potentials) predicts rich wetting phenomena
at walls, see e.g. references \cite{Eva83a,Kro85}. This is related to the underlying bulk equation
of state which can be derived from the mean--field functional: it is a reasonable
zeroth--order approximation to the true equation of state. The drawback of
the simplest models of the type discussed in references \cite{Eva83a,Kro85} 
consists in their complete failure to describe realistically fluid correlation functions.
This is due  to the inadequate treatment of the short--range repulsive forces. 
By now the treatment of short--ranged correlations within hard sphere density functionals has reached 
a very satisfactory level of precision and internal consistency;
for landmark developments see references \cite{Tar85,Ros89,Rot02}. By contrast, 
the treatment of the attractive
tails in the interatomic potentials has remained on the mean--field level
which leaves the equation of state treated in zeroth order approximation.

In order to relate density functional results in the presence of a wall, say, to quasi--exact
results as obtained from simulations, a number of methods have been proposed. 
First, one can observe that away from the critical region, density functional results
accurately reproduce simulation results if the mean--field coexistence curve is appropriately
scaled \cite{Swo89}. 
Second, close to exact correlation functions obtained from integral equation methods
may be used as input in density functionals, see references 
\cite{Cho01,Gil03,Tan03}. 
These amendments generally improve the agreement between 
density profiles obtained from theory and simulation. However, by introducing these ad--hoc 
modifications some key properties of the basic functionals (hard sphere functional 
plus mean field treatment of attractions)
are destroyed such as e.g. the compliance with the results of exact sum rules. Additionally, if this
modified class of density functionals is applied to reproduce the bulk fluid correlation
functions (this so--called test particle limit is obtained by choosing  
the fluid interatomic potential as the external potential) 
there is, in general, no agreement
between the integral equation input and the density functional output.

This requirement of test particle consistency was the starting point
for the density functional theory developed in \cite{Ros93}
which tries to combine standard density functional and integral equation
approaches more consistently. By making  a Taylor expansion
of the free energy functional around bulk densities the standard bulk integral
equations are derived and the bridge function in the bulk is determined via
a density functional for a suitably chosen reference system of hard spheres, 
thereby providing the necessary closure relation. The latter resembles the RHNC closure.
Results for the bulk structure of the one--component plasma \cite{Ros96}
and for a few state points of the Lennard--Jones fluid \cite{Ros98} show a similar 
level of agreement (if not better) with simulation data as RHNC results.

In this paper we will develop this idea, the reference functional approach, 
in more detail, generalizing the method
to an arbitrary inhomogeneous situation and also analyzing bulk fluid
and wall--fluid correlations using the mixture description.
We will see that by introducing the reference functional 
insertion free energies (i.e., the chemical potential in the bulk)
are obtained quite naturally. For the wall--fluid problem treated in the bulk mixture
approach this implies that the insertion free energy of the wall is explicitly
given; using it we can show that in the case of a hard wall complete drying is predicted.

The paper is structured as follows: In section \ref{sec:theory} we consider a fluid
mixture in an arbitrary inhomogeneous equilibrium state  and derive a general closure 
for the corresponding integral equations using the concept of the reference functional.
This permits us to derive expressions for the insertion free energies 
corresponding to the change in grand potential when inserting the  mixture 
particles into the inhomogeneous equilibrium state. 
In section \ref{sec:application}
we apply the approach to a mixture of a Lennard--Jones fluid and hard
particles. In the limit of infinite radius of the hard particles (upon which the
particles become hard walls) and infinitely small
hard particle density we solve the equations for the  fluid and wall--fluid  correlation 
functions. For the bulk fluid, we find very good agreement with simulation 
for both the virial and internal
energy equation of state. The liquid--vapour coexistence curve is obtained
by ($i$) the usual Maxwell construction and ($ii$) the requirement of complete drying at
the hard wall. 
We discuss the form of 
the wall--fluid bridge function required to account for complete drying. Furthermore,
we consider the consistency of our theory by analyzing two sum rules appropriate to the density distributions at a
hard wall. We discuss  the relation of the reference functional method to
commonly used density functional mean field models.   
Section \ref{sec:summary} contains a summary and perspectives for further research.

\section{Theory}
\label{sec:theory}

We consider a liquid mixture which contains $n$ components. The interaction potential between
atoms of species $i$ and $j$ is given by $u^{ij}(r)$ with $r=|\vect r_i - \vect r_j|$ denoting
the distance between the two atoms $(i,j=1\dots n)$. The $i^{\rm th}$ element of the vector 
$\boldsymbol \rho(\vect r)= \{\rho_1(\vect r),\dots,\rho_n(\vect r) \}$  describes the  
density distribution of species $i$ and likewise the $i^{\rm th}$ element of the vector
$\vect V(\vect r)=\{V_1(\vect r),\dots,V_n(\vect r) \}$ denotes the external potential
acting on an atom of species $i$. In equilibrium, there exists a unique correspondence 
between the external potential $\vect V(\vect r)$ and the density distribution which we denote
by $\boldsymbol \rho_{\vect V}$ to highlight this correspondence. As a consequence, 
there exists a unique free--energy functional \cite{Eva79}
\bea
  {\cal F}[\boldsymbol \rho] &=& {\cal F}^{\rm id} [\boldsymbol \rho] +
  {\cal F}^{\rm ex}  [\boldsymbol \rho] \\
  \beta {\cal F}^{\rm id} [\boldsymbol \rho] &=& \sum_i \int d\vect r\; \rho_i(\vect r)
   \left( \log (\rho_i(\vect r)\Lambda_i^3) -1 \right)
\eea
which is usually split into the exactly known ideal part (containing the
species--specific de--Broglie wavelength $\Lambda_i$) and the excess contribution.
The grand free energy functional is defined by
\bea
 \label{eq:omegadef}
  \Omega[\boldsymbol \rho] &=& {\cal F}[\boldsymbol \rho] - \int d\vect r \;
  ({\boldsymbol \mu}^{\rm id}(\rho_{i,0}) + \boldsymbol \mu^{\rm ex} (\rho_{i,0}) -
  \vect V(\vect r) ) \cdot \boldsymbol \rho (\vect r)
\eea
with $\boldsymbol \mu^{\rm id}(\rho_{i,0})$ denoting the chemical potential for an ideal mixture and 
$\boldsymbol \mu^{\rm ex}(\rho_{i,0})$ its excess (over ideal) for the asymptotic densities
$\boldsymbol \rho_0 = \{ \rho_{1,0},\dots, \rho_{n,0} \}$. The asymptotic densities correspond
to the homogeneous density distribution for $\vect V=0$. The components of the ideal chemical
potential are given by
$\beta \mu^{\rm id}_i = \log (\rho_{i,0}\Lambda_i^3$) and 
$\beta = 1/(k_B T)$ is the inverse 
temperature with Boltzmann's constant $k_B$.
The equilibrium density distribution follows from
\bea
 \label{eq:omegamin}
   \left. \frac{\delta \Omega}{\delta \boldsymbol \rho (\vect r)} \right|_{\boldsymbol \rho 
  = \boldsymbol \rho_{\vect V}} &=& 0 \;.  
\eea

The free energy functional generates the hierarchy of (inhomogeneous) 
direct correlation functions by
\bea
 \label{eq:c_hier}
 \beta  \left. \frac{\delta^n {\cal F}^{\rm ex}}
  {\delta \rho_{i_1}(\vect r_1) \dots \delta \rho_{i_n}(\vect r_n)}
   \right|_{\boldsymbol \rho = \boldsymbol \rho_{\vect V} }
 = - c^{(n),i_1\dots i_n}(\vect r_1, \dots, \vect r_n; \boldsymbol \rho_{\vect V} ) \,.
\eea
The first member of the hierarchy is determined through equation (\ref{eq:omegamin}):
\bea
 \label{eq:c1}
 c^{(1),i} (\vect r;\boldsymbol \rho_{\vect V}) &=& \log\frac{\rho_{{\vect V},i}(\vect r)}
 {\rho_{0,i}} - \beta \mu^{\rm ex}_i({\boldsymbol \rho}_0) + \beta V_{i}(\vect r) \;.
\eea
We are now interested in the change of the density distribution $\boldsymbol \rho_{\vect V}$ generated
through a perturbation corresponding to the external potential 
$\vect V' = \vect V + \Delta \vect V$. To this end we perform a functional Taylor
expansion of the unknown excess free energy in the function variable $\Delta\boldsymbol\rho(\vect r)
=\boldsymbol\rho(\vect r)-\boldsymbol \rho_{\vect V}(\vect r)$ and define:
\bea
  \label{eq:fsplit}
  {\cal F}^{\rm ex} = {\cal F}^{\rm HNC} + {\cal F}^{\rm B}\;.
\eea
Here, ${\cal F}^{\rm HNC}$ (the meaning of the label will become clear shortly) 
contains terms up to second order in $\Delta\boldsymbol\rho$ and all terms beyond
second order are subsumed in ${\cal F}^{\rm B}$. By virtue of definition (\ref{eq:c_hier}),
\bea
\beta {\cal F}^{\rm HNC}&=& \beta A^{\rm ex} ({\boldsymbol \rho}_{\vect V})
  - \sum_i \int d\vect r \; c^{(1),i} (\vect r;{\boldsymbol \rho}_{\vect V})
   \Delta \rho_i(\vect r) - \nonumber \\
 \label{eq:fhnc}
 &&  \frac{1}{2} \sum_{ij} \int d\vect r \;d\vect r'\; c^{(2),ij}(\vect r,\vect r';
  {\boldsymbol \rho}_{\vect V}) \Delta \rho_i(\vect r) \Delta \rho_j(\vect r')\;,
\eea
where $ A^{\rm ex}({\boldsymbol \rho}_{\vect V})$ is the excess free energy of the equilibrium state
corresponding to the external potential $\vect V$. Then the perturbed density distribution
$\boldsymbol \rho_{\vect V'}$ corresponding to the external potential $\vect V +\Delta\vect V$
is according to equations (\ref{eq:omegamin}) and (\ref{eq:c1})
\bea
  \label{eq:rhov1}
  \log \frac{\rho_{\vect V',i}(\vect r)}{\rho_{{\vect V},i}(\vect r)} + \beta \Delta V_i  &=&
  \sum_k \int d\vect r'\; c^{(2),ik}(\vect r,\vect r'; {\boldsymbol \rho}_{\vect V})
 \Delta \rho_{k} (\vect r') - \nonumber \\
    && \beta \left. \frac{\delta {\cal F}^{\rm B}}{\delta \rho_i(\vect r)}
  \right|_{\boldsymbol \rho=\boldsymbol \rho_{\vect V'}} \;.
\eea
It may appear that we have not gained very much from this formal manipulation: the density profile $\boldsymbol \rho_{\vect V'}$
is just expressed through the unknown density profile $\boldsymbol \rho_{\vect V}$, the
corresponding unknown inhomogeneous direct correlation function $c^{(2),ij}(\vect r,\vect r'; 
{\boldsymbol \rho}_{\vect V})$ and the unknown  functional ${\cal F}^{\rm B}$. Therefore we go
to the {\em test particle limit}, i.e. we define the perturbation  to $\vect V$
by fixing one atom
of species $j$ at position $\vect r_0$, thus $\Delta V_i (\vect r) = u^{ij}(\vect r - \vect r_0)$.  
For this particular choice of the perturbing potential, the density profiles are connected 
to the inhomogeneous pair correlation functions $g^{ij}$ and $h^{ij}=g^{ij}-1$ through
\bea
  \label{eq:gdef}
  \left. \rho_{\vect V',i}(\vect r) \right|_{V'_i(\vect r)=V_i(\vect r)+u^{ij}(\vect r - \vect r_0)} &= & 
   \rho_{\vect V,i}(\vect r) \; g^{ij}(\vect r,\vect r_0; {\boldsymbol \rho}_{\vect V}) \;, \\
  \label{eq:hdef}
 \left. \Delta \rho_{\vect V',i}(\vect r) \right|_{V'_i(\vect r)=V_i(\vect r)+u^{ij}(\vect r - \vect r_0)} &= &
  \rho_{\vect V,i}(\vect r) \; h^{ij}(\vect r,\vect r_0; {\boldsymbol \rho}_{\vect V})  \; .
\eea
The inhomogeneous pair correlation functions are  linked to the direct correlation functions via
the inhomogeneous Ornstein--Zernike relation
\bea
  h^{ij}(\vect r,\vect r_0; {\boldsymbol \rho}_{\vect V}) - 
  c^{(2),ij}(\vect r,\vect r_0; {\boldsymbol \rho}_{\vect V}) &=&
  \sum_k \int d\vect r'\; \rho_{{\vect V},k}(\vect r')  
  c^{(2),ik}(\vect r,\vect r'; {\boldsymbol \rho}_{\vect V} ) \cdot \nonumber \\
  \label{eq:inhOZ}
  && \qquad \qquad h^{kj}(\vect r',\vect r_0; {\boldsymbol \rho}_{\vect V}) \; .
\eea 
Inserting equations (\ref{eq:gdef})--(\ref{eq:inhOZ}) into equation (\ref{eq:rhov1})
we recover the general closure relation for the inhomogeneous correlation functions
\bea
 \label{eq:inh_cl}
 \log g^{ij}(\vect r,\vect r_0; {\boldsymbol \rho}_{\vect V}) + \beta u^{ij}(\vect r - \vect r_0)
 &=& h^{ij}(\vect r,\vect r_0; {\boldsymbol \rho}_{\vect V}) -
  c^{(2),ij}(\vect r,\vect r_0; {\boldsymbol \rho}_{\vect V}) - \nonumber \\
  \label{eq:ie_inh} 
  & &\beta \left. \frac{\delta {\cal F}^{\rm B}}{\delta \rho_i(\vect r)}
  \right|_{\rho_i=\rho_{\vect V,i}\; g^{ij}}\;,
\eea
which permits us to identify the density derivative of ${\cal F}^{\rm B}$ with the inhomogeneous
bridge function:
\bea
  \label{eq:bdef}
  b^{ij}(\vect r,\vect r_0; {\boldsymbol \rho}_{\vect V}) = 
    \beta \left. \frac{\delta {\cal F}^{\rm B}}{\delta \rho_i(\vect r)}
  \right|_{\rho_i(\vect r)=\rho_{\vect V,i}(\vect r)\; 
   g^{ij}(\vect r,\vect r_0; {\boldsymbol \rho}_{\vect V})} \;.
\eea
So far all is exact, and  at this point we recognize that upon setting ${\cal F}^{\rm B}=0$ we recover
the inhomogeneous HNC equations. Thus the HNC equations are generated by the second--order
functional for the excess free energy, ${\cal F^{\rm HNC}}$, {\em in the test particle limit}.
All higher orders in the density expansion of ${\cal F^{\rm ex}}$ contribute to the bridge
function.

The unknown density profile ${\boldsymbol \rho}_{\vect V}(\vect r)$ is connected to the
correlation functions through the exact YBG equations:
\bea
 \label{eq:YBG_h}
 \nabla \log \rho_{{\vect V},i}(\vect r) &=& \beta\,\nabla V_{i} (\vect r) +
  \beta\,\sum_j \int d\vect r'\; h^{ij}(\vect r, \vect r';{\boldsymbol \rho}_{\vect V}) \;
 \rho_{{\vect V},j} (\vect r') \; \nabla  V_{j} (\vect r')\;.
\eea
Using the inhomogeneous Ornstein--Zernike relation, equation (\ref{eq:inhOZ}), the last equation
can be transformed into one connecting $\rho_{\vect V,i }$ and $c^{(2),ij}$:
\bea
 \label{eq:YBG_c}
 \nabla \log \rho_{{\vect V},i}(\vect r) &=& \beta\,\nabla V_{i} (\vect r) +
 \sum_j \int d\vect r'\;c^{(2),ij}(\vect r, \vect r';{\boldsymbol \rho}_{\vect V}) 
  \nabla \rho_{{\vect V},j}(\vect r') \; .
\eea
Alternatively, this equation follows directly by taking the gradient in equation
(\ref{eq:c1}).
Thus we see that upon specification of a suitable model for ${\cal F}^{\rm B}$, we have a closed    
system of equations for $h^{ij}$, $c^{(2),ij}$ and $\rho_{\vect V,i }$ consisting of
equations (\ref{eq:inhOZ}), (\ref{eq:ie_inh}), and (\ref{eq:YBG_h}) 
or equations (\ref{eq:inhOZ}), (\ref{eq:ie_inh}), and (\ref{eq:YBG_c}). This set of equations
is standard in the theory of classical liquids and is usuallly derived using
diagrammatic expansions \cite{Ste64} or functional methods \cite{Per64}.

The excess free energy pertaining to the external potential
$\vect V$, $A^{\rm ex}({\boldsymbol \rho}_{\vect V})$, and the 
corresponding  one--body direct correlation
function $c^{(1),i} (\vect r;\boldsymbol \rho_{\vect V})$ do not appear in this system of equations. 
The latter can be determined using the Potential Distribution Theorem \cite{Hen83}
which states that $-c^{(1),i}/\beta$ is equivalent to the insertion
free energy, i.e. the chemical potential, of a particle of species $i$ into the 
inhomogeneous system defined by $\vect V$.
By definition, this insertion free energy is the difference in grand potential with
the test particle fixed at $\vect r_0$, say, and the grand potential without the test particle:
\bea
  \label{eq:c1def}
  \beta \mu_i^{\rm ex}(\vect r_0; \boldsymbol \rho_{\vect V}) & \equiv &  \beta\left.\Omega[{\boldsymbol \rho}]
  \right|_{\boldsymbol \rho = \boldsymbol \rho_{\vect V'}} - 
    \beta\left.\Omega[{\boldsymbol \rho}]\right|_{\boldsymbol \rho = \boldsymbol \rho_{\vect V}}
   = - c^{(1),i} (\vect r_0;\boldsymbol \rho_{\vect V})
\eea
where $V'_j(\vect r)=V_j(\vect r)+u^{ji}(\vect r-\vect r_0)$.
Explicit evaluation yields 
(using the definition of $\Omega$, equation (\ref{eq:omegadef}), the expressions
for the excess free energy, equations (\ref{eq:fsplit}) and (\ref{eq:fhnc}), and the definitions for
the inhomogeneous correlation functions in equations (\ref{eq:gdef}) and (\ref{eq:hdef})) 
\bea
 \fl
 -c^{(1),i} (\vect r_0) &=& \sum_j \int d\vect r \;
    \rho_{\vect V,j}(\vect r) \; g^{ji}(\vect r, \vect r_0) 
    \left( \log g^{ji}(\vect r, \vect r_0) + \beta u^{ji}(\vect r-\vect r_0) \right ) - \nonumber \\
 \fl
   &&  \frac{1}{2}\sum_{jk} \int d\vect r \int d\vect r' \; \rho_{\vect V,j}(\vect r)
      \rho_{\vect V,k}(\vect r') \; c^{(2),jk} (\vect r,\vect r') h^{ji}(\vect r ,\vect r_0)
    h^{ki}(\vect r' ,\vect r_0) - \nonumber \\
 \fl
  &&  \sum_j \int d\vect r \; \rho_{\vect V,j}(\vect r) \; h^{ji}(\vect r, \vect r_0) +
   \beta \left.{\cal F}^{\rm B}[{\boldsymbol \rho}]\right|_{
    \rho_j(\vect r)=\rho_{\vect V,j}\; g^{ji}(\vect r,\vect r_0)}  \;.
\eea
For compactness, we have suppressed the variable ``$\boldsymbol \rho_{\vect V}$'' in all correlation
functions which indicates that these pertain to the inhomogeneous density distribution
defined by $\vect V$. We note that in deriving the last equation, $A^{\rm ex}$ and 
$c^{(1),i}$ have dropped out on the right hand side. To simplify this expression, we use
the general closure relation (\ref{eq:inh_cl}), the inhomogeneous Ornstein--Zernike equation 
(\ref{eq:inhOZ}) and the definition (\ref{eq:bdef}):
\bea
 \label{eq:c1_ins} 
 -c^{(1),i} (\vect r_0) &=& -c^{(1),i}_{\rm HNC}(\vect r_0)-
     \sum_j \int d\vect r\; \rho_{{\vect V},j}(\vect r) \;g^{ji}(\vect r,\vect r_0)\;
      b^{ji}(\vect r,\vect r_0) + \nonumber \\
   && \beta \left.{\cal F}^{\rm B}[{\boldsymbol \rho}]\right|_{
    \rho_j(\vect r)=\rho_{\vect V,j}\; g^{ji}(\vect r,\vect r_0)} \;.
\eea 
This result, which is the main formal result of our paper, constitutes 
the generalization of the well--known HNC result for the chemical potential in inhomogeneous
systems\footnote{See reference \cite{Kje89} for a derivation using the well--known 
charging method.} to closures with non--vanishing bridge functions. 
Here, $-c^{(1),i}_{\rm HNC}(\vect r_0)/\beta$ describes the 
position--dependent HNC insertion free energy for 
${\cal F}^{\rm B}=0$:  
\bea
\fl \label{eq:inh_muhnc}
 -c^{(1),i}_{\rm HNC}(\vect r_0) = \sum_j \int d\vect r\;
   \rho_{{\vect V},j}(\vect r)\left( \frac{1}{2} h^{ji}(\vect r,\vect r_0)\left[
  h^{ji}(\vect r,\vect r_0)-c^{(2),ji}(\vect r,\vect r_0)\right] - c^{(2),ji}(\vect r,\vect r_0)
 \right) \;.\; \nonumber \\
\eea
Note that $c^{(1),i}$ is determined by the density distribution
${\boldsymbol \rho}_{\vect V}$ through equation (\ref{eq:c1}) which imposes a consistency
constraint on all subsequent approximations to ${\cal F}^{\rm B}$.
Considering  the limit $|\vect r_0| \to \infty$, we may assume that the external potential
vanishes, $\vect V(\vect r_0) \to 0$, thus ${\boldsymbol \rho}_{\vect V} \to
{\boldsymbol \rho}_{0}$ and the inhomogeneous correlation functions become
the bulk correlation functions,
 $g^{ji}(\vect r,\vect r_0) \to g^{ji}(\vect r-\vect r_0)$. Then equations 
(\ref{eq:c1}) and (\ref{eq:c1_ins}) imply an equation for the excess chemical
potential of species $i$ {\em in the bulk}:
\bea
 \label{eq:muex} \fl
 \beta \mu_i^{\rm ex}({\boldsymbol \rho}_{0}) &=& 
   \beta \mu_i^{\rm ex, HNC}({\boldsymbol \rho}_{0}) -
   \sum_j \rho_{0,j} \int d\vect r\; g^{ji}(\vect r)\;
      b^{ji}(\vect r) + %\nonumber \\
    \beta \left.{\cal F}^{\rm B}[{\boldsymbol \rho}]\right|_{
    \rho_j(\vect r)=\rho_{0,j}\; g^{ji}(\vect r)} \;, \\
 \fl
 \beta \mu_i^{\rm ex, HNC}({\boldsymbol \rho}_{0}) &=& \sum_j \rho_{0,j} \int d\vect r\;
   \left( \frac{1}{2} h^{ji}(\vect r)\left[
  h^{ji}(\vect r)-c^{(2),ji}(\vect r)\right] - c^{(2),ji}(\vect r) \right) \;.
\eea    
The same equation is derived if $\vect V=0$ is assumed from the outset, i.e. if
the functional expansion of the excess free energy is performed around the bulk
state described by ${\boldsymbol \rho}_{0}$.

In general, the excess free energy functional beyond second order,
the bridge functional  ${\cal F}^{\rm B}$, is not known.
However, we may approximate ${\cal F}^{\rm B}$ by a density functional for a reference system in the following
manner:
\bea
 \label{eq:fbref}
 {\cal F}^{\rm B}[\boldsymbol \rho] \approx {\cal F}^{\rm B,ref}[\boldsymbol \rho] &=& 
    {\cal F}^{\rm ref}[\boldsymbol \rho]- 
          {\cal F}^{\rm HNC,ref}[\boldsymbol \rho]\;, %\\ 
\eea
where the second order HNC contribution is given by equation (\ref{eq:fhnc}):
\bea
\beta {\cal F}^{\rm HNC, ref}[\boldsymbol \rho]&=& \beta A^{\rm ex,ref} ({\boldsymbol \rho}_{\vect V})
  - \sum_i \int d\vect r \; c^{(1),i}_{\rm ref} (\vect r;{\boldsymbol \rho}_{\vect V})
   \Delta \rho_i(\vect r) - \nonumber \\
 \label{eq:fhnc_ref}
 &&  \frac{1}{2} \sum_{ij} \int d\vect r \;d\vect r'\; c^{(2),ij}_{\rm ref}(\vect r,\vect r';
  {\boldsymbol \rho}_{\vect V}) \Delta \rho_i(\vect r) \Delta \rho_j(\vect r')\;,
\eea 
For an  explicitly given functional ${\cal F}^{\rm ref}$, the direct correlation functions
$c^{(1),i}_{\rm ref}$ and $c^{(2),ij}_{\rm ref}$ are obtained through equation (\ref{eq:c_hier}). 
The reference functional approximation implies that all direct correlations beyond
second order are well approximated by the corresponding direct correlations of the reference system.
Equivalently, it states that the contributions to the excess free energy beyond
those of the reference system
are of second order in the density difference $\Delta \boldsymbol \rho$,
\bea
  \Delta A &=& {\cal F}^{\rm ex}[\boldsymbol \rho] - {\cal F}^{\rm ex,ref}[\boldsymbol \rho] 
          = {\cal F}^{\rm HNC}[\boldsymbol \rho] - {\cal F}^{\rm HNC,ref}[\boldsymbol \rho] \; \\
      &=& {\rm const.} - \frac{1}{\beta}
    \sum_i \int d\vect r \; (c^{(1),i}-c^{(1),i}_{\rm ref}) (\vect r;{\boldsymbol \rho}_{\vect V})
   \Delta \rho_i(\vect r) - \nonumber \\
 \label{eq:mf_gen}
&&  \frac{1}{2\beta} \sum_{ij} \int d\vect r \;d\vect r'\; (c^{(2),ij}-c^{(2),ij}_{\rm ref})
   (\vect r,\vect r'; {\boldsymbol \rho}_{\vect V}) \Delta \rho_i(\vect r) \Delta \rho_j(\vect r')\;.
\eea 
Thus the reference functional approximation defines a kind of a generalized mean--field 
approximation in treating a general fluid beyond a given reference system where the coefficient
in the quadratic term, $c^{(2),ij}-c^{(2),ij}_{\rm ref}$, is determined self--consistently.

In practice, there are only three reference systems for which we possess reasonably accurate
density functionals. They are all related to hard--sphere systems where we can use
geometric arguments for their construction. These encompass 
$(i)$ functionals of Rosenfeld type for hard--sphere mixtures (see e.g. references \cite{Ros89,Rot02}),
$(ii)$ a functional for the Asakura--Oosawa colloid--polymer mixture \cite{Bra03}, and 
$(iii)$ a recently proposed functional for non--additive  hard--sphere mixtures 
\cite{Sch04}. The functionals of type $(i)$ have been tested in a variety of inhomogeneous
situations and can be regarded as very robust. The functional $(ii)$ 
has been applied to the structure of free interfaces and near a hard wall, predicting
a wealth of interesting phenomena \cite{Bra03}. A drawback is that the polymers
are treated only linearly in their density, consequently correlations 
from the functional are deficient if they are sensitive to higher orders in the polymer 
density. 
The functional $(iii)$ has not been tested yet
for inhomogeneous situations. It contains the functional for the Asakura--Oosawa
model as a limiting case, so similar merits and problems may be expected there.

Summarizing the literature which is concerned with the reference functional idea, we note
that the formalism presented above has been   formulated previously only for 
functionals of type $(i)$ and for functional expansions around bulk densities ${\boldsymbol \rho}_0$,
i.e. $\vect V =0$.
In fact, the reference functional approximation 
as expressed by equation (\ref{eq:fbref})  was
introduced in reference \cite{Ros93} and used to describe the one--component plasma 
with high accuracy and its similarity to the reference HNC equations was also pointed out. 
Later on, the numerical efficacy of this idea was illustrated  
for a few state points of the one--component Lennard--Jones (LJ) fluid \cite{Ros98}
and of binary mixtures with their size asymmetry being not too large \cite{Kah96}. 
Comparison with simulation results for both the correlation functions and
thermodynamic properties shows very good agreement. 

Although the hard sphere functionals of type $(i)$ are very robust, they are not exact.
For example, the bulk direct correlation function obtained from the functionals by direct
functional differentiation, equation (\ref{eq:c_hier}), represent good approximations 
for its behaviour inside the hard core but predict it to be zero outside. 
In order to improve upon the accuracy of the correlation functions, the reference functional 
method may be applied to the hard sphere reference system itself. As a result,
for one--component hard spheres the bridge function for the bulk
integral equations generated by using the White Bear functional of reference 
\cite{Rot02} appears to be the most accurate one known up to now.
Applied to binary hard--sphere mixtures and size asymmetries $>10$, the bulk integral
equation closure shows deficiencies in calculating the distribution function  
between the big spheres \cite{Oet04}. This may be related  to the inadequacy of the 
functional Taylor expansion around {\em bulk} densities for large size asymmetries
or to the inadequacy of the functionals themselves (as manifest in a low--density
analysis of the third--order direct correlation function $c^{(3)}$ \cite{Cue02}).

The advantage of the present formalism is that for a general, inhomogeneous density distribution
${\boldsymbol \rho}_{\vect V}$ it is possible to determine directly  
insertion free energies using equation (\ref{eq:c1_ins}). Of much interest is e.g. obtaining
the insertion free energy for colloids at fluid interfaces which determines their adsorption
behaviour at the interface.
However, a numerical solution of equations (\ref{eq:inhOZ})--(\ref{eq:YBG_h}) requires 
large numerical efforts. Functional minimization in two or three dimensions  is complicated
by the nature of the three types of functionals  mentioned above; these
contain $\delta$ and $\theta$--like distributions.

In this paper, we confine ourselves to an illustration of  the approach for a simple one--component fluid 
and for $\vect V =0$. On the one hand, the solution to the homogeneous Ornstein--Zernike
equation with the integral equation closure defined by
equation (\ref{eq:inh_cl}) gives the pressure $p(\rho)$ and the chemical 
potential $\mu(\rho)$ for a fixed temperature
via the virial equation of state and equation (\ref{eq:muex}), respectively. This
permits the determination of the liquid--vapour coexistence curve via a standard Maxwell
construction as is usually done in integral equation theories. On the other hand, by considering
a binary mixture of the fluid with hard spherical cavities in the limit of vanishing cavity density and
infinite cavity radius we can derive a wall--particle integral equation. The onset 
of complete drying at the planar hard wall (the cavity with infinite radius) 
permits an independent determination of the coexistence curve.
We examine the degree of consistency between these two different routes.

\section{Application: Equation of state for a simple fluid and drying at a hard wall}
\label{sec:application}

We consider a binary mixture of cut--off and shifted Lennard--Jones particles
(species 1) with hard cavities of radius $R_{\rm c}$ (species 2). Let
\bea
  u_{\rm LJ}(r) &=& 4\epsilon\left[ (r/\sigma)^{-6} - (r/\sigma)^{-12} \right]
\eea
then the interaction potentials in the mixture are given by
\bea 
  u^{11}(r) &=& \left\{ \begin{array}{ll} 
   u_{\rm LJ}(r) -  u_{\rm LJ}(r_c) &\quad (r \le r_c)  \\
           0 &\quad (r>r_c) 
  \end{array} \right.
   ,\; \\
  \exp(-\beta u^{12}(r))=\exp(-\beta u^{21}(r)) &=& \theta( R_{\rm c} - r)\;, \\
  \exp(-\beta u^{22}(r)) &=& \theta( 2R_{\rm c} - r) \; .
\eea
We apply the formalism from the last chapter with ${\boldsymbol \rho}_{\vect V}=
\{\rho_{1,0}, \rho_{2, 0}\}$ and for the reference system we  use the original 
Rosenfeld functional \cite{Ros89}
which contains two unknown hard--sphere diameters $d_1$ and $d_2$. In the limit
$\rho_{2,0} \to 0$ ($\rho_{1,0} \equiv \rho_0$) the equations for $h^{11}\equiv h$ and 
$c^{(2),11}\equiv c^{(2)}$ (fluid
correlations) and $h^{12}$ (cavity--fluid correlation) decouple.
The integral equation system for the LJ fluid (cf. equations  (\ref{eq:inhOZ}) and
 (\ref{eq:inh_cl})) is given by: 
\bea
 \label{eq:hlj}
 \log g(\vect r) + \beta u^{11}(\vect r)  &=&h(\vect r) -  c^{(2)}(\vect r) -
  \left.\beta \frac{\delta {\cal F}^{\rm B,ref}}
   {\delta \rho(\vect r)}\right|_{\rho(\vect r)=\rho_{0}\; g(\vect r)} \;,  \\
 \label{eq:clj}
  h(\vect r) -  c^{(2)}(\vect r) &=& \rho_{0}\; h \ast c^{(2)}(\vect r) \\
  \label{eq:conv}
 &&  \left( h \ast c^{(2)}  (\vect r) = \int d\vect r' c^{(2)}(\vect r-\vect r')\;
  h(\vect r') \right) \;,
\eea
with $g(\vect r) \equiv g(r)$ etc.
Following the reference functional assumption, ${\cal F}^{\rm B}$ has been replaced
by the corresponding functional for the reference system, cf. equation
(\ref{eq:fbref}), and the fluid bridge function  is given by
\bea
  b(\vect r) &=& \left.\beta \frac{\delta {\cal F}^{\rm B,ref}}
   {\delta \rho(\vect r)}\right|_{\rho(\vect r)=\rho_{0}\; g(\vect r)} \;.
\eea
Using the Ornstein--Zernike equation in the limit $\rho_{2,0} \to 0$ in equation
(\ref{eq:inh_cl}) yields a single equation for the cavity--particle correlation which,
however, needs as input the solution for $c^{(2)}$, the pair direct fluid correlation 
function:
\bea
 \label{eq:h12}
 \log g^{12}(\vect r) + \beta u^{12}(\vect r)  &=&  \rho_{0}\; h^{12} \ast c^{(2)}(\vect r)
  - \left.\beta \frac{\delta {\cal F}^{\rm B,ref}}
   {\delta \rho(\vect r)}\right|_{\rho(\vect r)=\rho_{0}\; g^{12}(\vect r)} \;.
\eea
In taking the planar hard wall limit $R_{\rm c} \to \infty$, we shift 
the origin of the coordinate system such that 
$\lim_{R_{\rm c} \to \infty} \exp(-\beta u^{12}(\vect r))=\theta(z)$. In this limit,  
the cavity--fluid correlation functions will be referred to as the 
wall--fluid correlation functions. 

Note that in the dilute limit $\rho_{2,0} \to 0$ both the system of equations 
(\ref{eq:hlj}), (\ref{eq:clj}) and equation (\ref{eq:h12}) can be derived 
directly from minimization
of the grand potential employing the excess free energy functional 
{\em of the one--component LJ fluid},
${\cal F}^{\rm HNC}[\rho(\vect r)] + {\cal F}^{\rm B,ref}[\rho(\vect r);d_1]$, and
using $u^{11}$ and $u^{12}$, respectively, as external potentials. Consequently,
the reference functional ${\cal F}^{\rm B,ref}$ is derived from the one--component
Rosenfeld functional and contains only one unknown parameter, the hard--sphere
diameter $d_1$.

Both equations (\ref{eq:hlj}) and (\ref{eq:h12}) are similar to the reference HNC
closure \cite{Lad83}. There, the bridge function is approximated directly by the bridge
function of the reference system whereas here the bridge function is obtained from
the reference system {\em free energy functional}. For state points near the triple 
point, both prescriptions agree with each other closely. However, 
in reference HNC the fluid bridge function remains short--ranged near the bulk 
critical point as does the wall--fluid bridge function near coexistence on the liquid 
side. Thus reference HNC omits effects of bulk criticality and of drying at the hard 
wall.
In the present scheme, the fluid bridge
function becomes long--ranged through its functional dependence on the 
fluid correlation function $g$ (which is long--ranged near the critical point)
and so does the wall--fluid bridge function near coexistence  through its dependence
on the wall--particle correlation function $g^{12}$.
 As will be seen, this does not lead to the correct bulk critical behaviour
for the fluid but drying at the hard wall is described very well.
   
For the determination of the unknown  hard--sphere diameter $d_1$ we propose
the following criterion,
\bea
 \label{eq:crit2}
  \frac{\partial}{\partial d_1} \left( {\cal F}^{\rm B,ref}[ \rho_0\: g(\vect r);d_1]
   - {\cal F}^{\rm B,ref}[ \rho_0\: g^{\rm ref}(\vect r);d_1] \right)
    \stackrel{!}{=} 0 \;,
\eea
which corresponds to extremizing the free energy difference between the fluid
and the reference system
with respect to $d_1$. The correlation function of the reference system $g^{\rm ref}$
is taken to be the Percus--Yevick correlation function consistent with
the Rosenfeld functional. This criterion is local in the bulk state space
$(\rho,T)$ and  turns out to be practicable and 
reliable, furthermore it is seen in the numerical results that in the
moderate to high density region it is equivalent to the
usual reference HNC criterion \cite{Lad83}.
However,
a deeper understanding of equation (\ref{eq:crit2}) with regard
to a unique specification of the reference system is absent at this point. 

For the numerical calculations, we use the cut--off $r_c = 4 \,\sigma$ for which 
an extensive body of simulation data for the equation of state exists \cite{Joh93}.

\subsection{Equation of state}

\begin{figure}[b]
\caption{\label{fig:eos} (colour online)
 Left panel: Virial equation of state at various temperatures, $p^\ast = p^\ast(\rho_0^\ast)$. 
 The pressure is calculated according to equation (\ref{eq:virial}).
  Right panel: Internal energy equation of state at various temperatures, 
 $\beta u =\beta u (\rho_0^\ast)$. 
 The excess internal energy is calculated according to equation (\ref{eq:internal}).
The symbols denote simulation data taken from reference \cite{Joh93}.}
 \vspace*{10mm}
 \begin{center}
   \epsfxsize=7.5cm
   \epsfbox{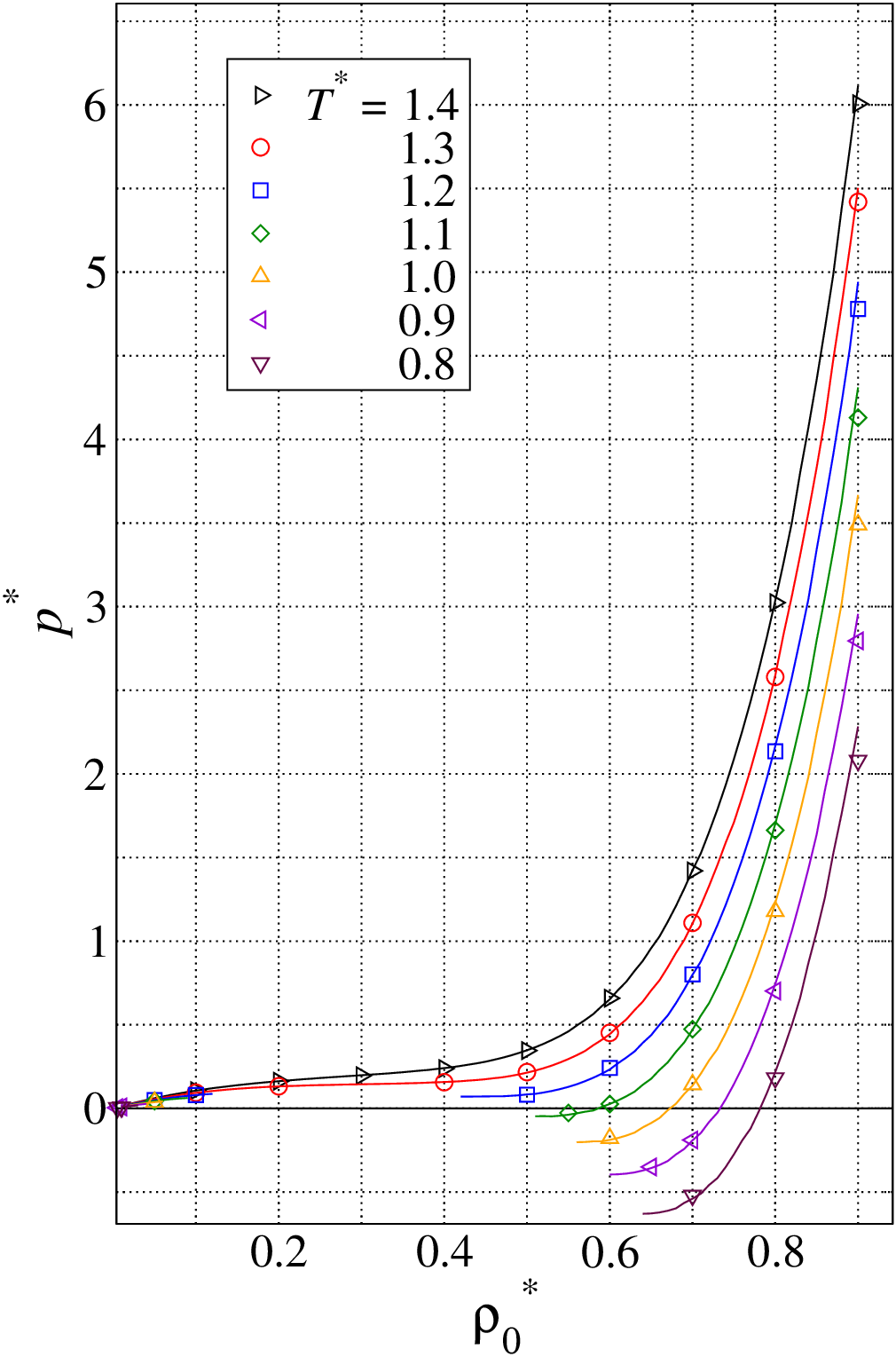} \hspace{3mm}
   \epsfxsize=7.5cm
   \epsfbox{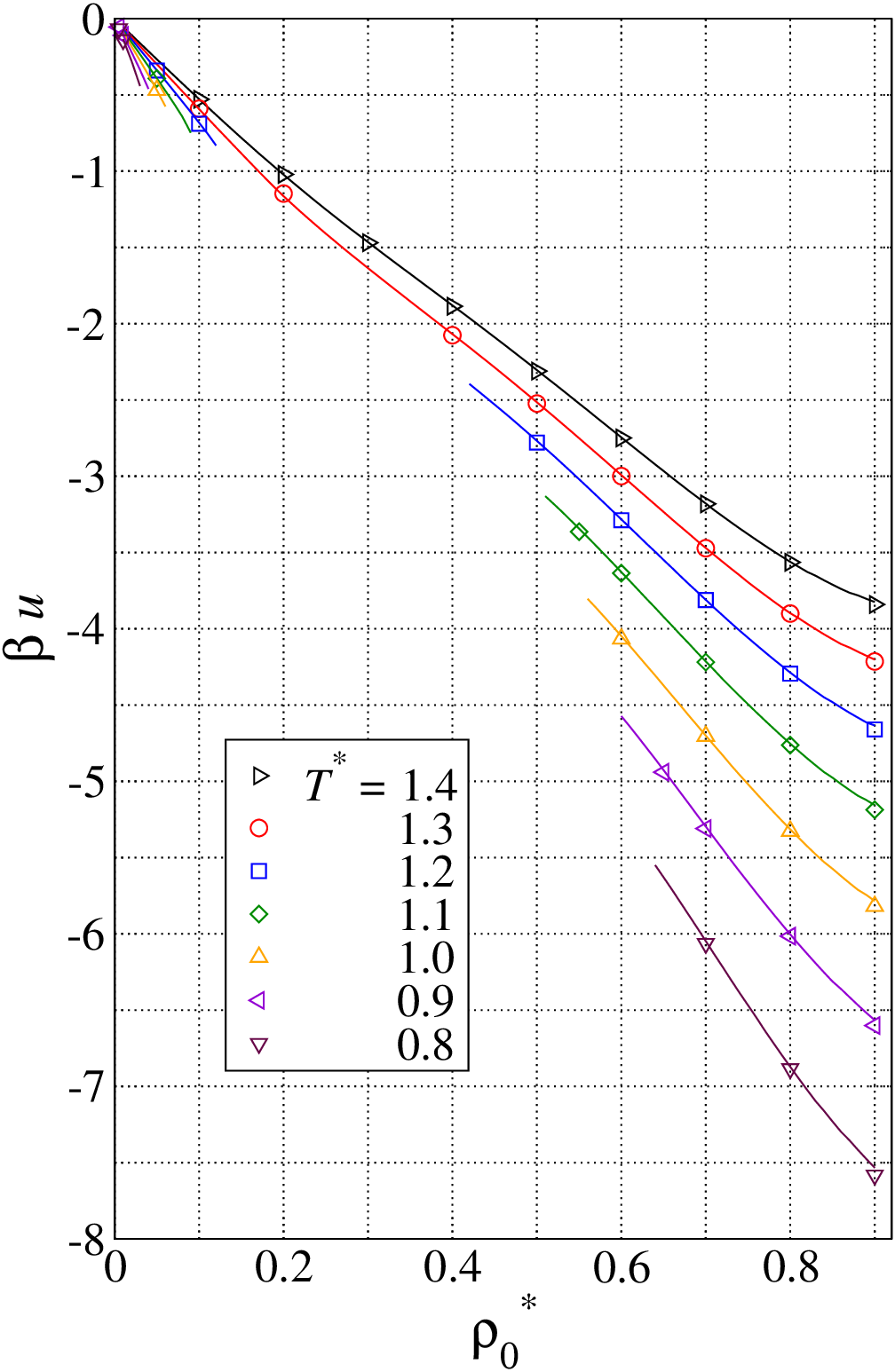}
 \end{center}
\end{figure}

The pressure $p$ and the excess (over ideal) internal energy per particle $u$ are determined 
from the fluid pair correlation function through the well--known equations:
\bea
  \label{eq:virial}
  \frac{\beta p}{\rho_0} &=& 1 - \frac{2}{3}\, \pi\, \rho_0\,  \beta \int_0^\infty
    dr\, r^3 \,g(r) \frac{d u^{11}(r)}{dr} \; , \\
  \label{eq:internal}
     u &=& 2\pi\, \rho_0\,  \int_0^\infty dr\, r^2\, g(r)\,  u^{11}(r) \; .
\eea
The usual consistency check of integral equation closures proceeds via the
compressibility relation
\bea
 \label{eq:com}
  \beta \frac{\partial p}{\partial \rho_0} &=& 1-4\pi \int_0^\infty dr\, r^2\,c^{(2)}(r) \;.
\eea
In principle, the excess Helmholtz free energy per particle $a^{\rm ex}$ can be determined
through integration of equation (\ref{eq:internal}) along isochores,
\bea
  \beta a^{\rm ex} &=& \int_0^\beta u(\beta')\,d\beta',
\eea
 and the excess chemical potential follows by differentiating on an isotherm
\bea
 \label{eq:mu_a}
  \mu^{\rm ex} &=& a^{\rm ex} + \rho_0 \frac{\partial a^{\rm ex}}{\partial \rho_0}\;. 
\eea  
However, in the reference functional approach the excess chemical potential is 
given directly by (see equation (\ref{eq:muex}))
\bea
  \label{eq:mubulk}
  \beta \mu^{\rm ex} &=& \beta \mu^{\rm ex, HNC} - 
   4\pi\,\rho_0 \int_0^\infty dr\,r^2\, b(r)\,g(r) +
    \beta {\cal F}^{\rm B,ref}[\rho_0\, g(r);d_1]\;, \\
  \beta \mu^{\rm ex, HNC} &=& 4\pi\,\rho_0 \int_0^\infty dr\,r^2\,  
  \left( \frac{1}{2}h(r)\left[ h(r)-c^{(2)}(r)\right] - c^{(2)}(r) \right) \;.
\eea
Consistency with the virial equation (\ref{eq:virial}) may be checked using the
isothermal Gibbs--Duhem relation
\bea
  \label{eq:GD}
  \frac{\partial p}{\partial \mu} = \rho_0\;, \qquad (\mu = \mu^{\rm id}+\mu^{\rm ex} )\; .
\eea

\begin{figure}
\caption{\label{fig:consistency} (colour online)
Testing the consistency of the equation of state at various temperatures. 
Upper panel: The ratio of the 
density derivatives of the pressure, calculated ($i$) using the
compressibility equation (\ref{eq:com}) and ($ii$) via the derivative of 
a polynomial fit
to the virial pressure, equation (\ref{eq:virial}). Lower panel: The ratio
of the chemical potential derivative of the virial pressure and the bulk density.
In order to perform the derivative, both the virial pressure, 
equation (\ref{eq:virial}), and the chemical potential,
equation (\ref{eq:mubulk}), were fitted to polynomials in the bulk density.
In both panels, the density coordinate of the filled circles
for the three lowest temperatures $T^\ast$  indicates the corresponding coexistence    
density of the liquid phase obtained from simulations \cite{Joh93}.
}
 \vspace*{12mm}
 \begin{center}
   \epsfxsize=12cm
   \epsfbox{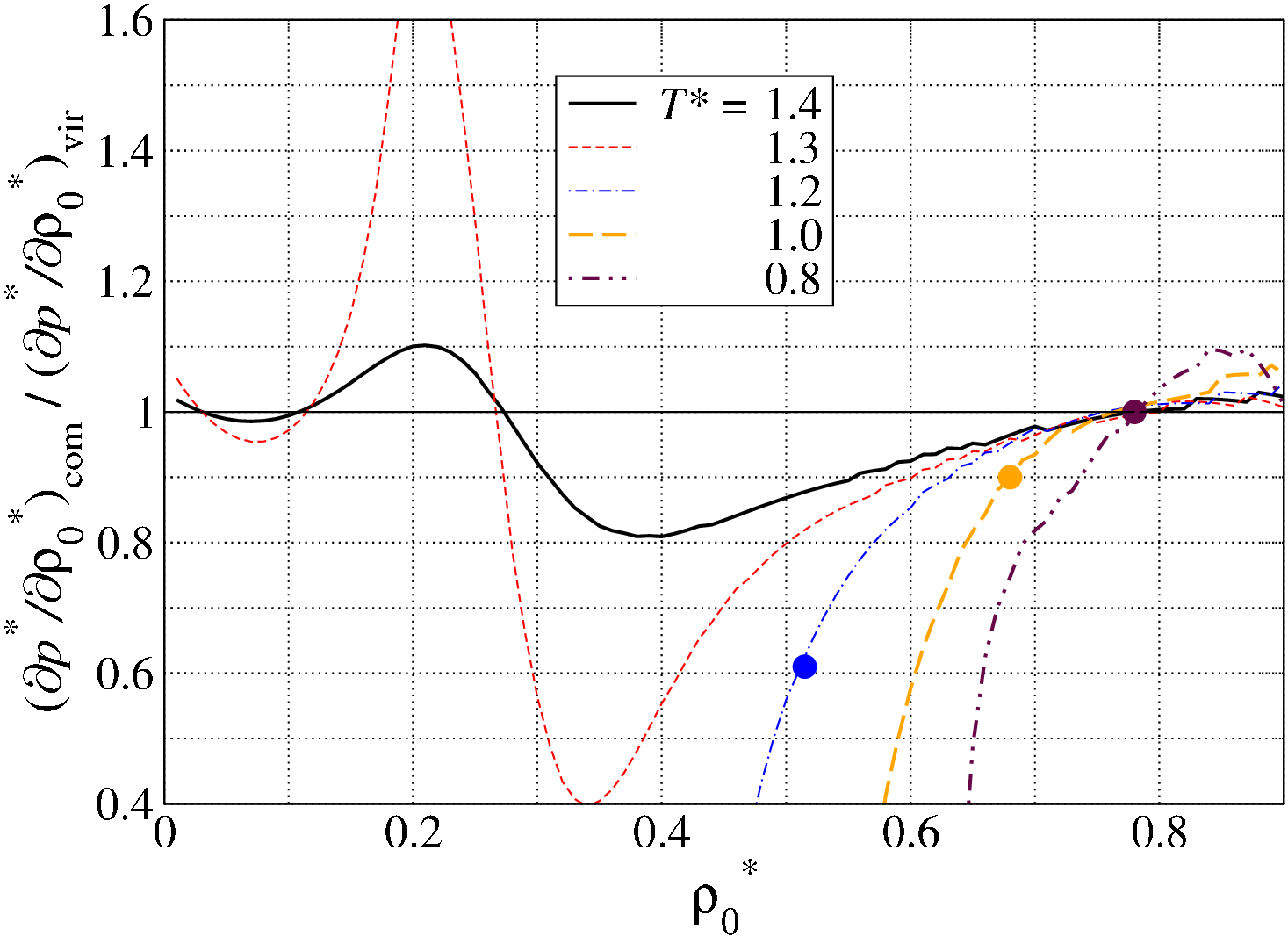}\\[2cm] 
   \epsfxsize=12cm
   \epsfbox{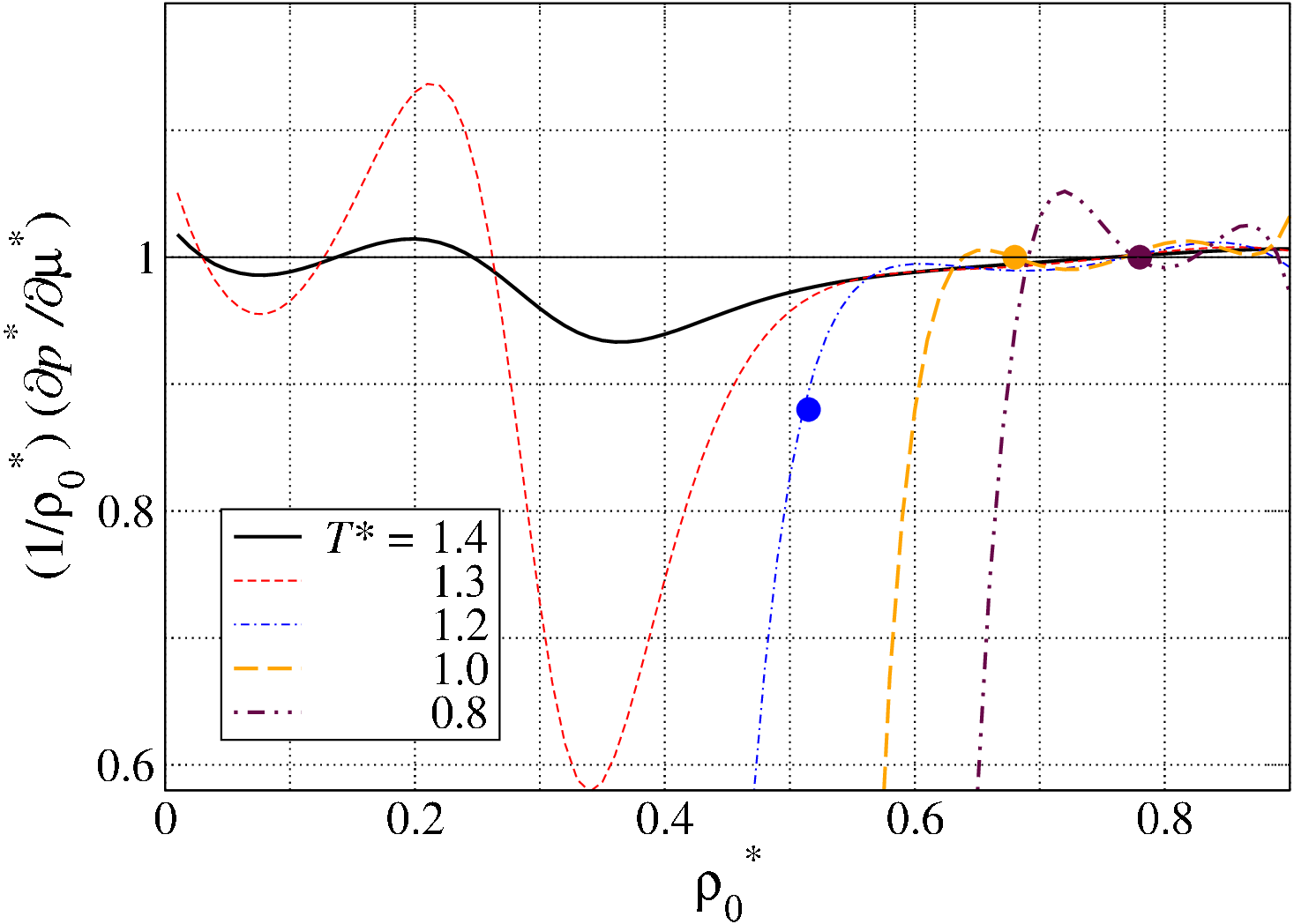}
 \end{center}
\end{figure}

From the simulations of reference \cite{Joh93} the critical temperature and density
of the cut--off and shifted LJ fluid
can be estimated as $T_{\rm c}^\ast \approx 1.25$ and 
$\rho_{\rm c}^\ast \approx 0.31$\footnote{Reduced units, denoted by an asterisk, 
are defined by setting $\epsilon=\sigma=1$.}. In Figure \ref{fig:eos}, we
present our results for $p^\ast(\rho_0^\ast)$ and $\beta u(\rho_0^\ast)$ for 
temperatures $T^\ast = 0.8 \dots 1.4$. We see that for all temperatures
and densities (both close to and away from the critical region) the pressure
and internal energy calculated from the reference functional  
match the simulation data very well. This indicates that
the pair correlation function $g(r)$ is very accurate within the 
range of $u^{11}(r)$ (see equations
(\ref{eq:virial}) and (\ref{eq:internal})). In Figure \ref{fig:consistency},
we check the consistency of the virial and the compressibility equation of state
(upper panel) and the Gibbs--Duhem relation (lower panel) 
with the chemical potential calculated
via the insertion method, equation (\ref{eq:mubulk}). Clearly, for the critical region,
roughly delineated by $T^\ast = 1.2 \dots 1.3$ and $\rho_0^\ast = 0.15 \dots 0.5$,
we observe inconsistencies between the virial and compressibility equation of state,
and, somewhat weaker, deviations from $(\partial p/\partial\mu)\,/\,\rho_0=1$. However, outside
the critical region and for stable bulk densities $\rho_0$, the ratio
of $\partial p/\partial\rho_0$ calculated via equation (\ref{eq:com}) 
and via the derivative of
equation (\ref{eq:virial}) indicates violations of at most 20\%, and the
Gibbs-Duhem relation is satisfied rather well. This will be important for the 
subsequent investigation of drying at the hard wall.

\subsection{Two routes to liquid--vapour coexistence and drying at the hard wall}

Integral equations usually give results for pair correlation functions in the
metastable as well as in the stable region, and therefore it is customary to obtain
 the coexistence curve
by requiring $\mu(\rho_{\rm l})=\mu(\rho_{\rm g})$ and 
$p(\rho_{\rm l})=p(\rho_{\rm g})$; this is equivalent
to employing a common tangent or Maxwell construction. The density of the
coexisting liquid is given by $\rho_{\rm l}$, while $\rho_{\rm g}$ denotes the
coexisting gas density. The pressure is given by equation (\ref{eq:virial})
and the chemical potential may be calculated by either equation (\ref{eq:mu_a})
or in our case equation (\ref{eq:mubulk}). 

However, one may avoid the 
problem of dealing with correlation functions in the metastable domain altogether
if one defines $\rho_{\rm l}$ as the bulk density for which complete drying
at a hard wall occurs at a given $T<T_c$. Correspondingly, $\rho_{\rm g}$ is 
then the density in
the bulk of the infinitely extended drying film.
 
In order to understand the physics of drying \cite{Hen85}, let us consider a liquid 
of density $\rho_0$ near
a hard wall at fixed temperature which is close to coexistence, expressed
by $\delta\mu=\mu(\rho_0)-\mu(\rho_{\rm l})>0$ and $\delta\mu$ small.  
Since the wall exerts no attractive forces on the fluid molecules, 
we would expect the fluid density near the wall to be gaslike as this is energetically
more favourable. Corroborating this argument, there exists a sum rule
relating the fluid density at the hard wall, $\rho_{\rm c}$, 
to the bulk pressure of the liquid,
\bea
 \rho_{\rm c} &=&\beta p(\mu(\rho))\; \nonumber \\
              &\approx & \beta( p(\mu(\rho_{\rm l})) + \rho_l \delta\mu) \;, 
  \nonumber \\
              & \approx& \rho_{\rm g} + \beta(\beta_1 \rho_{\rm g}^2/2 + 
            \rho_l \delta\mu )\;,
 \label{eq:prule1}
\eea
(see also equation (\ref{eq:prule}) below). Since the
second virial coefficient of a simple liquid, $\beta_1$, is negative, this sum rule tells us
that $\rho_{\rm c}<\rho_{\rm g}$ for $\delta\mu \to 0$. Therefore the density profile
must pass from a value smaller than the coexisting gas density at the wall
to the liquid density in the bulk. Since we can safely assume (for $T<T_c$) that
the transition from $\rho_{\rm g}$ to $\rho_{\rm l}$ happens within a few
molecular diameters we are led to the hypothesis that upon $\delta\mu \to 0$
a gas layer forms between the hard wall and the bulk liquid   
whose width $l$ goes to infinity as $\delta\mu\to 0$ {\em for all $T<T_c$}
-- complete drying occurs.
More formally, the emergence of the gas film may be treated using a coarse--grained 
interface Hamiltonian  which in  mean--field
approximation predicts a slow divergence of
the gas film width upon approaching coexistence: $l \propto - \ln \delta \mu$.
For a recent general analysis of liquids with short--ranged attractions near spherical
hard walls, see reference \cite{Eva04}. 
%Furthermore we know that fluctuation effects
%do not alter the mean--field predictions for complete drying at all $T<T_c$
%\cite{Die88}. 

The coexisting densities in the hard wall route are obtained by inspecting the grand 
potential 
\bea
 \label{eq:om_wall}
\Omega[\rho] &=& {\cal F}^{\rm id}[\rho(\vect r)] + {\cal F}^{\rm HNC}[\rho(\vect r)] + 
   {\cal F}^{\rm B,ref}[\rho(\vect r)] - \nonumber \\
  & & \int dr( \mu(\rho_{\rm l}) - u^{12}(\vect r) ) \rho(\vect r)\; ,
\eea
with $u^{12}(\vect r) \to u^{12}(z)$, the planar hard wall potential.
The excess free energy functional ${\cal F}^{\rm HNC}+{\cal F}^{\rm B,ref}$ is given  
by the  Taylor expansion around the liquid bulk density $\rho_0=\rho_{\rm l}$.
When examining the stability conditions of an infinitely extended drying film
between the wall and the infinitely extended bulk liquid, we can neglect 
the hard wall potential $u^{12}$ and the grand potential functional
in equation (\ref{eq:om_wall}) simply becomes the bulk grand potential in a Taylor
expansion around $\rho_{\rm l}$.
As a first condition, the grand potential of the drying film with density
$\rho_{\rm g}$ must be stationary
\bea
 \left. \frac{\delta \Omega}{\delta \rho(\vect r)}\right|_{\rho(\vect r)=\rho_{\rm g}} 
  &\stackrel{!}{=}& 0\;.
\eea
 Let $V$ denote the system volume. We introduce the bulk densities of the
grand potential, $\omega(\rho)=\Omega(\rho)/V$ and   of the free energy,
$f(\rho)={\cal F}(\rho)/V$. The second condition is that
the drying film can only coexist with the liquid if the grand potential densities
for the two bulk densities $\rho_{\rm g}$ and $\rho_{\rm l}$ are equal:
\bea
   \omega(\rho_{\rm g})-\omega(\rho_{\rm l})
   &\stackrel{!}{=}& 0\;.
\eea
We insert the grand potential of equation (\ref{eq:om_wall})
into the above two conditions, evaluate them at the required coexisting bulk densities 
and using the abbreviations $\Delta \mu=
\mu(\rho_{\rm g})- \mu(\rho_{\rm l})$, $\Delta\omega=\omega(\rho_{\rm g})- 
\omega(\rho_{\rm l})$ we obtain: 
\bea
 0 \stackrel{!}{=} \beta \Delta \mu &=& 
  \beta\mu^{\rm ex,ref}(\rho_{\rm g};d_{1,{\rm l}}) -
  \beta\mu^{\rm ex,ref}(\rho_{\rm l};d_{1,{\rm l}}) + \nonumber \\ 
 \label{eq:deltamu}
 &&\left[\tilde c^{(2)}(0;\rho_{\rm l})- 
  \tilde c^{(2),{\rm ref}}(0;
  d_{1,{\rm l}})\right]\;(\rho_{\rm g}-\rho_{\rm l})+ 
  \log\frac{\rho_{\rm l}}{\rho_{\rm g}} \;, \\
 0 \stackrel{!}{=} \Delta \omega &=&  f^{\rm ex,ref}(\rho_{\rm l};d_{1,{\rm l}})- 
   f^{\rm ex,ref}(\rho_{\rm g};d_{1,{\rm l}}) +\frac{1}{\beta}
  (\rho_{\rm l}-\rho_{\rm g}) - \nonumber \\
 && \left[\rho_{\rm l} \,  \mu^{\rm ex,ref}(\rho_{\rm l};d_{1,{\rm l}}) -
    \rho_{\rm g}\, \mu^{\rm ex,ref}(\rho_{\rm g};d_{1,{\rm l}})\right] + \nonumber \\
 \label{eq:deltaom}
  && \frac{1}{2\beta} \left[\tilde c^{(2)}(0;\rho_{\rm l})-
  \tilde c^{(2),{\rm ref}}(0;
  d_{1,{\rm l}})\right]\;(\rho^2_{\rm l}-\rho^2_{\rm g})
\eea 
Here, $d_{1,{\rm l}}=d_1(\rho_{\rm l})$ is the reference hard--sphere diameter
which is determined by condition (\ref{eq:crit2}) and
$\tilde c^{(2)}(k)$ is the Fourier transform of $c^{(2)}(r)$. 
The system of equations (\ref{eq:deltamu}) and  (\ref{eq:deltaom})
can be solved for $\rho_{\rm l}$ and $\rho_{\rm g}$ along with the simultaneous
solution to the integral equations (\ref{eq:hlj}) and (\ref{eq:clj}) 
for the liquid (which is needed to obtain $\tilde c^{(2)}(0;\rho_{\rm l})$).

\begin{figure}
\caption{\label{fig:coexistence} (colour online)
Liquid--gas coexistence for the cut--off and shifted LJ fluid.
The curve is derived from an empirical fit to the free energy of the 
full LJ fluid, using mean--field corrections to account for the 
cutoff and shift in the potential \cite{Joh93}. The squares are the coexistence points
derived by the Maxwell construction using equations (\ref{eq:virial}) and 
(\ref{eq:mubulk}). The diamonds are the coexistence points obtained by considering
 complete drying at the hard wall expressed through equations  (\ref{eq:deltamu}) 
and (\ref{eq:deltaom}). 
}
 \vspace*{12mm}
 \begin{center}
   \epsfxsize=12cm
   \epsfbox{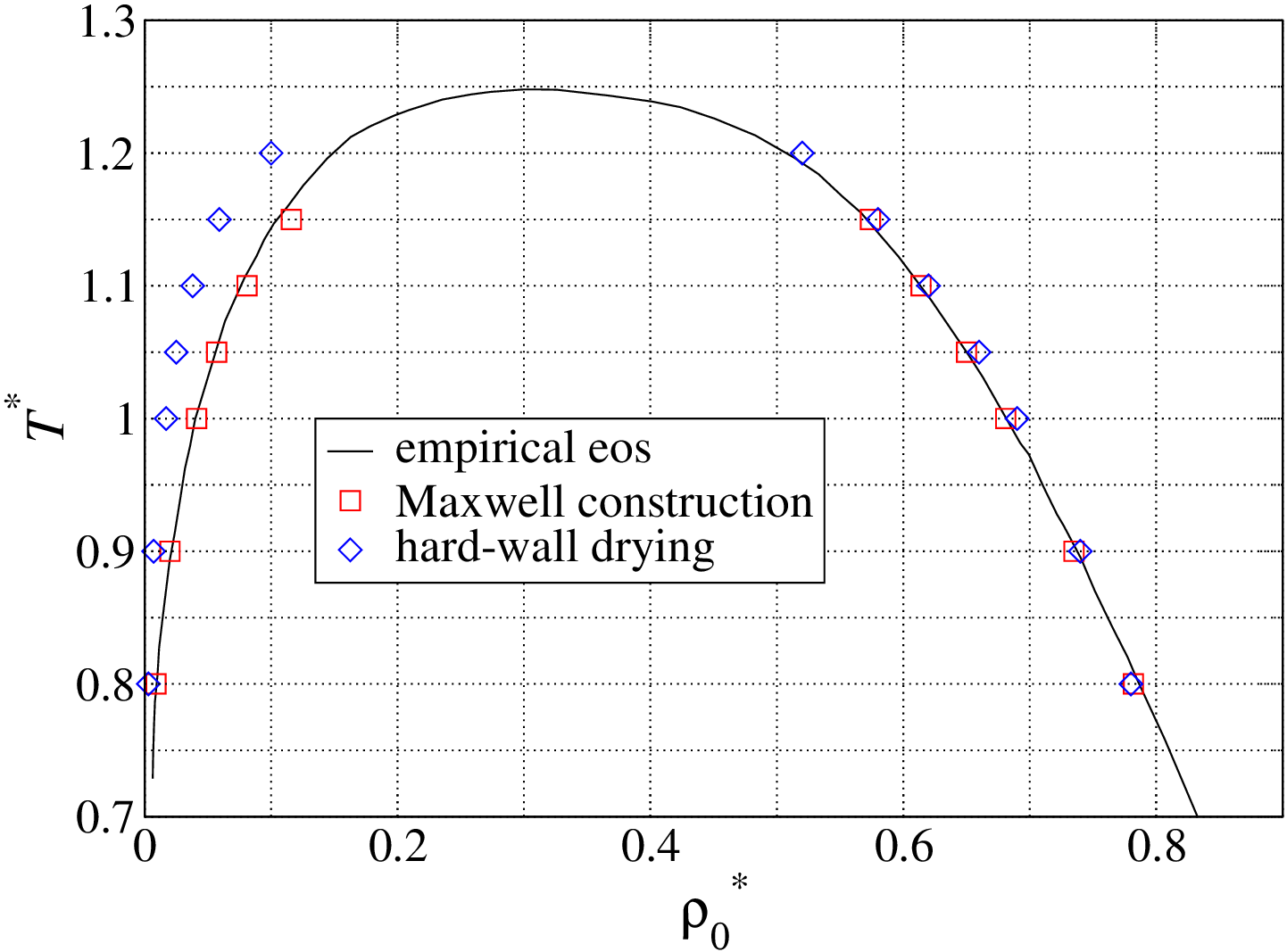}
 \end{center}
\end{figure}

In Figure \ref{fig:coexistence} we compare the coexistence curve from the standard Maxwell
construction (using equations (\ref{eq:virial}) and (\ref{eq:mubulk})) and
that from hard--wall drying with a curve fitted to simulation data
 from reference \cite{Joh93}
which was obtained from an equation of state. 
In agreement with our findings concerning the accuracy of the Gibbs--Duhem
equation (cf. Figure (\ref{fig:consistency})), the Maxwell construction gives
an excellent description of the coexistence curve away from the critical region. Closer
to the critical point, we are not able to find  a solution using the Maxwell 
construction. This can be
traced back to the violation of the Gibbs--Duhem equation. Hard--wall
drying yields coexisting liquid densities with similar accuracy but there are
clear deviations from simulations
for the coexisting gas densities are clearly visible. The reason for this lies in
the error in
extrapolating the bulk grand potential density, equation (\ref{eq:deltaom}),
to low densities, using only bulk properties of the reference system and
the pair direct correlation function of the coexisting liquid state point.
Nevertheless, the fair agreement between the results from the Maxwell construction 
and from hard--wall drying
is most encouraging, and, to our knowledge, has not been achieved before
within integral equation approaches.

\begin{figure} 
\caption{\label{fig:hardwall} (colour online) Surface tension (top panel) and adsorption density
(bottom panel) of the cut--off and shifted LJ fluid at a hard wall at temperature
$T^\ast=1.0$. Symbols are integral equation results and curves correspond to a fit
near coexistence
to the functional form given in equations (\ref{eq:g_mf}) and (\ref{eq:G_mf}),
setting $\gamma_{\rm wg}(\mu)$ to a constant and neglecting $\Gamma_{\rm wg}(\mu)$. 
From the fit we extract the bulk correlation length $\xi \approx 0.45\,\sigma$ and
$a \approx 0.89\, k_BT/\sigma^3$.
The range of the chemical 
potential covered by the integral equation results corresponds to the bulk
densities $\rho_0^\ast=0.686\dots 0.9$.
}
 \vspace*{12mm}
 \begin{center}
   \epsfxsize=12cm
   \epsfbox{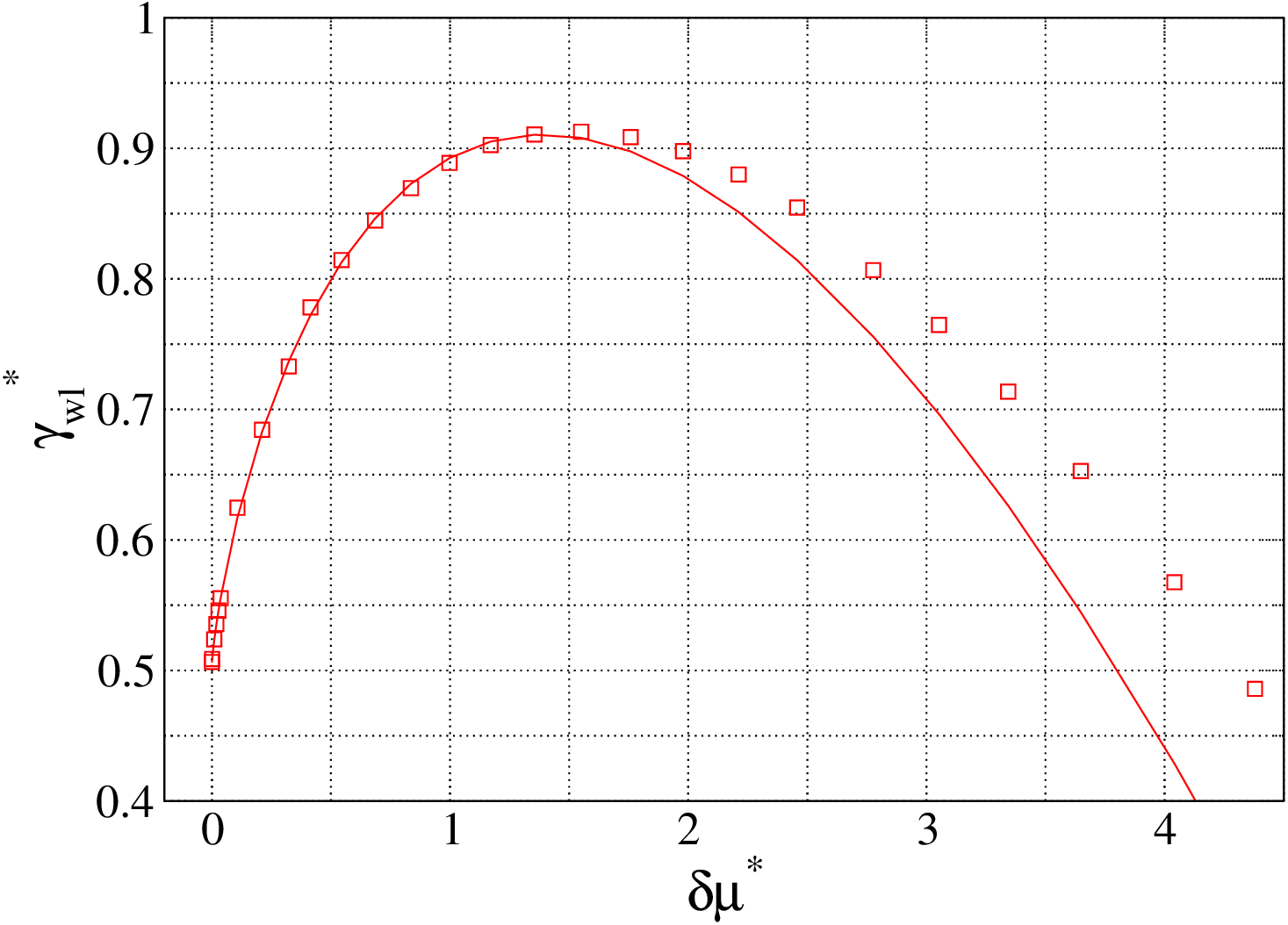}\\[2cm]
   \epsfxsize=12cm
   \epsfbox{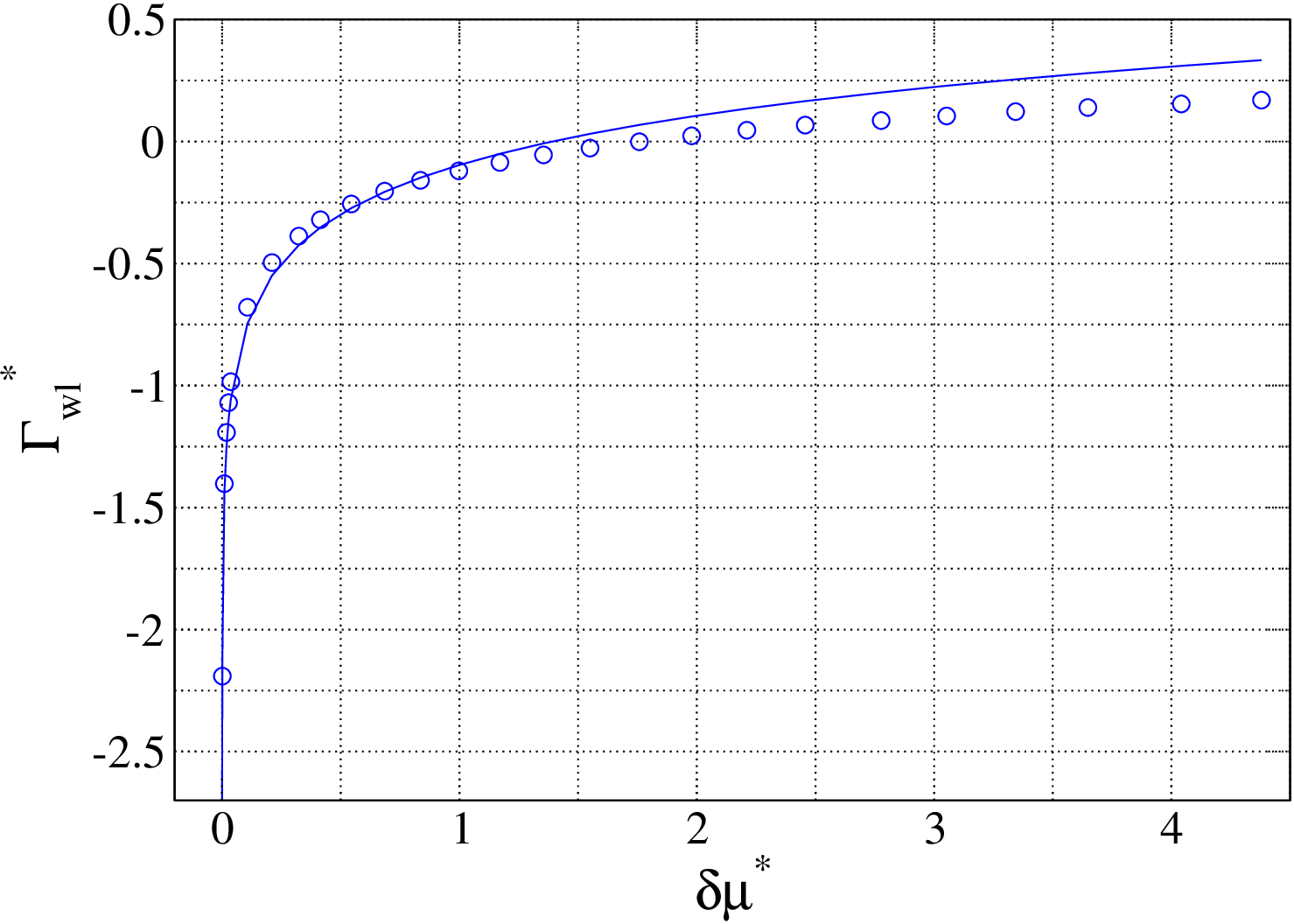}
 \end{center}
\end{figure}

Upon approaching the coexisting liquid density from above on an isotherm, the
wall--liquid adsorption density  $\Gamma_{\rm wl}$ diverges and the wall--liquid 
surface tension $\gamma_{\rm wl}$ approaches the sum of the 
liquid--gas surface tension $\gamma_{\rm gl}^\infty$
and the  wall--gas surface tension $\gamma_{\rm wg}$.
A mean field analysis of an effective interface Hamiltonian for
this situation gives \cite{Die88}:
\bea
 \label{eq:g_mf}
 \gamma_{\rm wl}(\mu) &=& \gamma_{\rm wg}(\mu) + \gamma_{\rm gl}^\infty +
 \xi\;\Delta\rho \;\delta\mu \left[
   \ln\left( \frac{a}{\Delta\rho\,\delta\mu}\right) +1  \right]\;, \\ 
 \label{eq:G_mf}
 \Gamma_{\rm wl}(\mu) &=& \Gamma_{\rm wg}(\mu) - \xi\;\Delta\rho \; 
  \ln\left( \frac{a}{\Delta\rho\,\delta\mu}\right)\;.
\eea
Here, $\mu=\mu(\rho_0)$ is the chemical potential in the bulk, far away from the 
wall ($\delta\mu=\mu-\mu(\rho_{\rm l}))$, $\Delta\rho=\rho_{\rm l}-\rho_{\rm g}$ 
is the difference of the coexisting densities  and $\xi$ 
is the decay length of the correlation function in the gas phase.
%is the gas phase bulk correlation length (the decay length of the gas tail of the planar gas--liquid interface). 
The effective parameter $a$ is of order $O(k_B T/\xi^3)$.
Fluctuations are expected to change the above behaviour only slightly; a treatment
within the interface Hamiltonian approach \cite{Die88} yields a renormalization of
the bulk correlation length: $\xi \to (1+\omega/2)\,\xi$ with $\omega=
k_B T/(4\pi\,\gamma_{\rm gl}^\infty\,\xi^2)$.

We have analyzed the surface tension and the adsorption density for the
isotherm $T^\ast = 1.0$. The adsorption density is given by 
\bea
 \label{eq:Gamdef}
 \Gamma_{\rm wl} = \rho_0 \int_0^\infty h^{12}(z) dz\;, 
\eea
while the surface tension $\gamma_{\rm wl}$ can be derived 
from the general formula (\ref{eq:c1def}).
Its explicit form is given in \ref{app:gamma}. In Figure \ref{fig:hardwall}
we show the surface tension and the adsorption density obtained from the
reference functional theory together with a fit to the mean--field prediction
given by equations (\ref{eq:g_mf}) and (\ref{eq:G_mf}). We observe that the 
latter describes the integral equation results not only close to the coexisting
density $\rho_{\rm l}^\ast \approx 0.686$ but also up to densities 
$\rho_0^\ast = 0.8$ ($\delta \mu^\ast \approx 1.5$) where we can observe a 
crossover of the adsorption density
from negative to positive values which is equivalent to a maximum in the
surface tension. This result is consistent with the idea  that the wall--fluid
correlations are derived from a generalized mean field functional;
see equation (\ref{eq:mf_gen}) and the discussion thereafter. 
Note that results very similar to those in  Figure \ref{fig:hardwall} were obtained
in one of the pioneering studies of hard--wall drying \cite{Hen85}, 
in simulations for a square--well fluid.

\begin{figure}
\caption{\label{fig:hardwall1} (colour online)
 Wall--liquid bridge functions from the  reference functional method (red lines)
and reference HNC (blue lines) at temperature $T^\ast =1.0$. Upper panel: Bulk
density $\rho_0^\ast=0.6859\approx \rho_{\rm l}$. Lower panel:  Bulk density 
$\rho_0^\ast=0.9$. The wall--fluid correlation function $g^{12}$ from the
reference functional method is given by the dashed line. For the
density close to coexistence, $g^{12}$ displays a thick drying film whose
extent is similar to that of $b^{12}$, i.e. about $5\,\sigma$. By contrast,
$b^{12}_{\rm RHNC}$ remains short--ranged.
}
 \vspace*{15mm}
 \begin{center}
   \epsfxsize=12cm
   \epsfbox{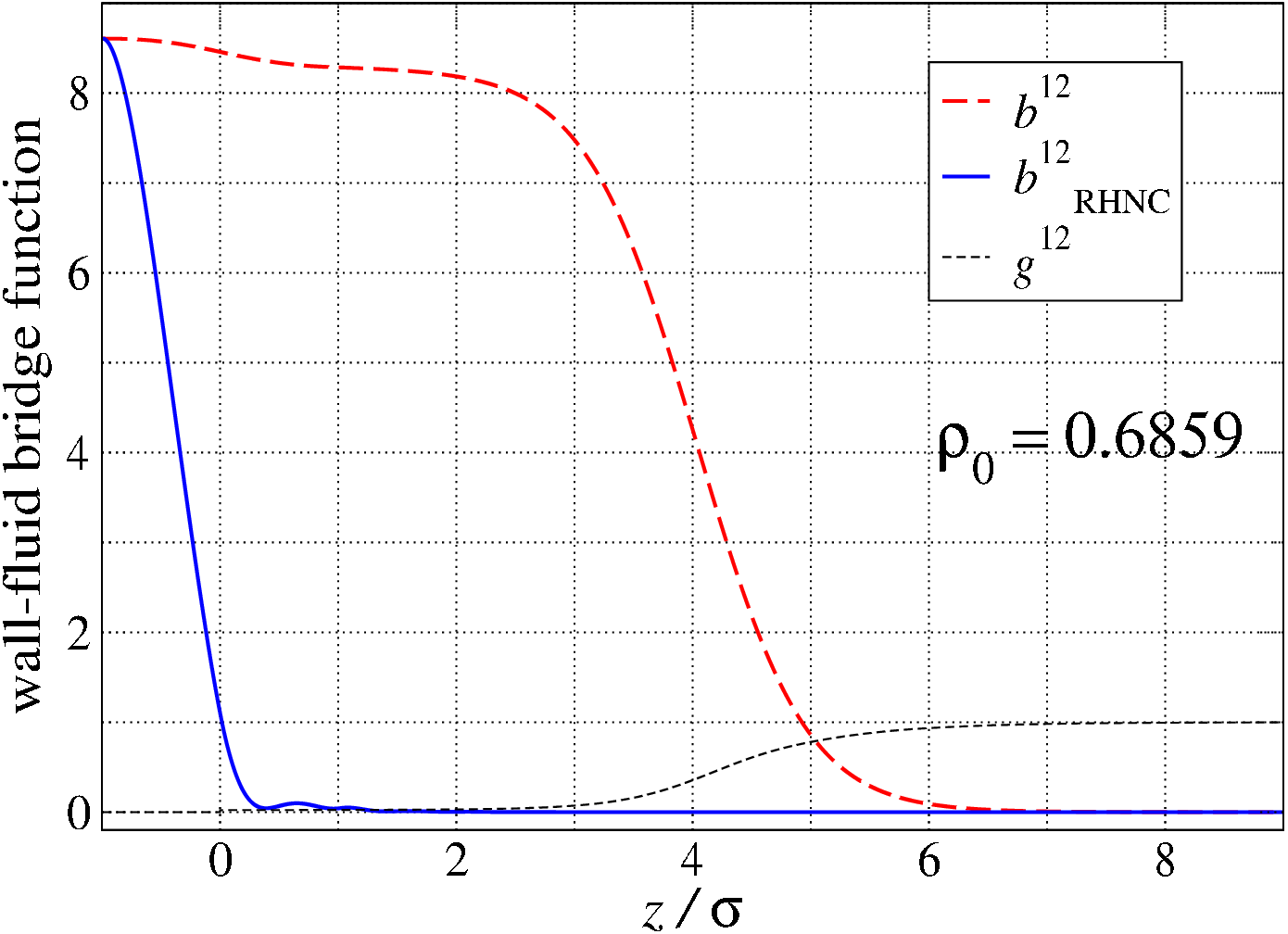}\\[2cm]
   \epsfxsize=12cm
   \epsfbox{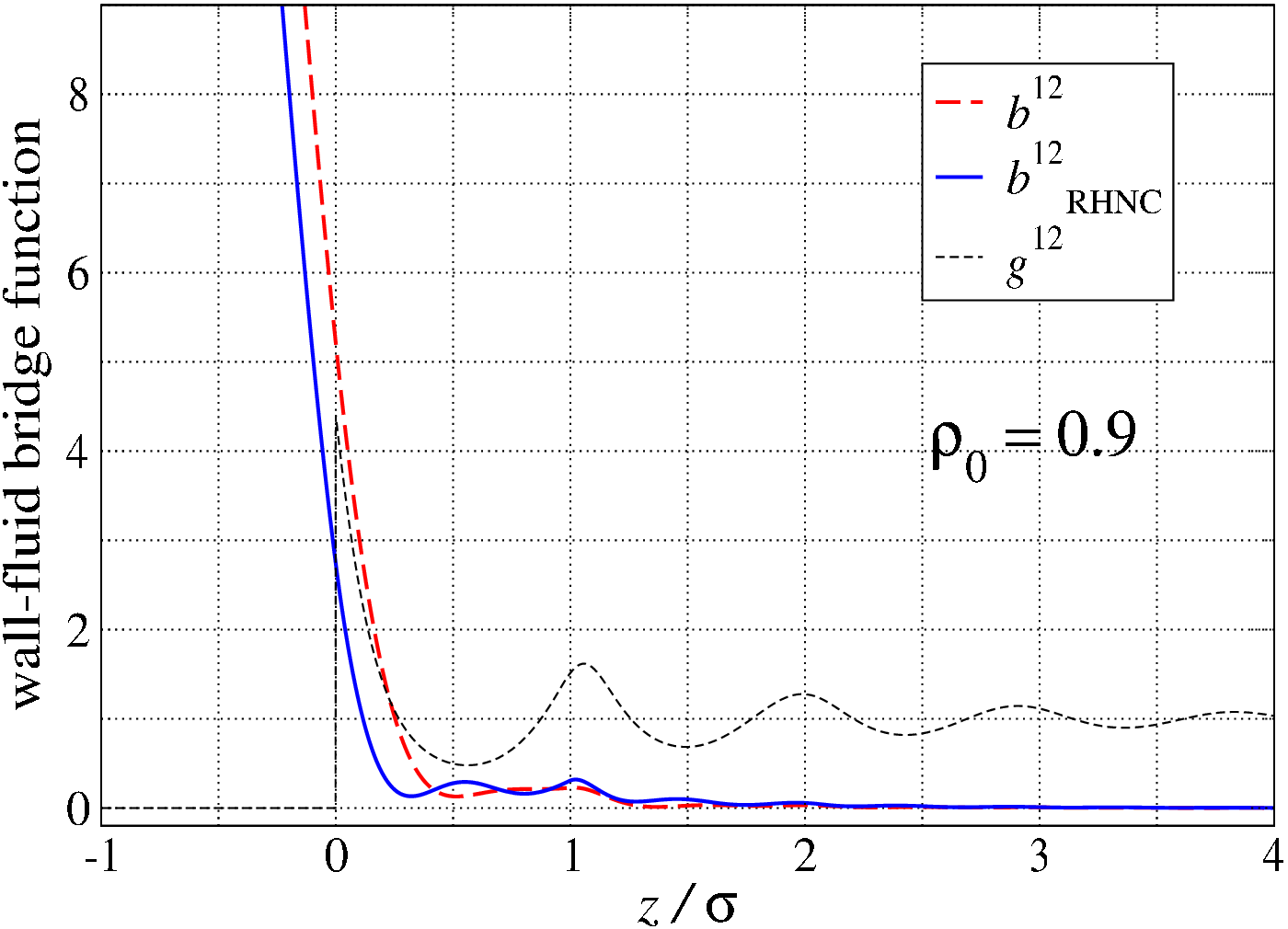}
 \end{center}
\end{figure}

Next, we compare the predictions for the wall--fluid 
bridge function $b^{12}(z)$ using the reference functional with those of  the reference HNC
equations. This comparison is shown in Figure \ref{fig:hardwall1} for a state point 
very close to coexistence ($\rho_0^\ast = 0.6859$) and a state point deep in
the stable liquid domain ($\rho_0^\ast = 0.9$). In reference HNC, the bridge function
is derived purely from the reference system (hard spheres with diameter $d_1$ at
the hard wall) and this bridge function is approximated very well by
\bea
  b^{12}_{\rm RHNC}(z) &=& \left.\beta \frac{\delta {\cal F}^{\rm B,ref}}
   {\delta \rho(\vect r)}\right|_{\rho(\vect r)=\rho_{0}\; g^{12,{\rm ref}}
   (z)} \;.
\eea
Here, $g^{12,{\rm ref}}$ is the wall--fluid correlation function for hard spheres
at the hard wall obtained by minimizing the Rosenfeld hard sphere functional in the
presence of the hard wall potential. For the higher bulk density 
(lower panel in Figure \ref{fig:hardwall1}) $b^{12}$
and $b^{12}_{\rm RHNC}$ show very similar, short--ranged behaviour 
whereas
for the density close to coexistence (upper panel in Figure \ref{fig:hardwall1})
$b^{12}$ acquires a long--ranged component of approximately the same extent as the
drying film (compare with the plot of $g^{12}$ in  Figure \ref{fig:hardwall1}).
Such a long--range behaviour is completely absent in $b^{12}_{\rm RHNC}$ 
which stays short--ranged and 
consequently the drying film formation is not captured by the reference HNC
equations.

\subsection{Comparison with the hydrostatic HNC approach}
\label{sec:hhnc}

At this point it is worthwhile to compare the present approach with the hydrostatic
HNC (HHNC) equations of references \cite{Zho90a,Zho90b} which is to the 
author's knowledge the only 
systematic attempt to account for wetting (drying) phenomena within integral equation
theories. Following reference \cite{Zho90a}, we imagine that an atom of species 2 
(originally our hard cavities) exerts a very weak and slowly varying potential on 
an atom of species 1. Let the slowly varying correlation functions between 
species 2 and species 1 be denoted by $h_{\rm s}$ and $g_{\rm s}=h_{\rm s}+1$.
Then the HHNC bridge function is
\bea
  b^{12}_{\rm HHNC}(\vect r) = \beta \mu^{\rm ex}(\rho_0\,g_{\rm s}(\vect r))- 
                \beta \mu^{\rm ex}(\rho_0) -
                     \rho_0 h_{\rm s}(\vect r) \; \frac{\partial(\beta \mu^{\rm ex})}
                      {\partial \rho}(\rho_0)\;.
\eea
Here, $\mu^{\rm ex}(\rho)$ is the excess chemical potential of the fluid (species 1) 
which is
supposed to be known exactly. This expression for the bridge function is
exact for slowly varying potentials (hydrodynamic limit). Furthermore we note
the thermodynamic identity
\bea
  \rho_0  \frac{(\beta \partial \mu^{\rm ex})} {\partial \rho}(\rho_0) =
  - 4\pi \int_0^\infty dr\, r^2\,c^{(2)}(r) = - \tilde c^{(2)}(0)
\eea
which follows from equations (\ref{eq:com}) and  (\ref{eq:GD}).
In order to apply this
approximation for the bridge function to potentials which vary sharply over molecular distances
(i.e., the hard wall potential or the fluid potential $u^{11}$) the 
authors of references \cite{Zho90a,Zho90b} 
introduce a normalized weight function
$w(\vect r)$ which smoothes the correlation functions:
\bea
 h_{\rm s}(\vect r)=w \ast h^{12}(\vect r), \qquad 
 g_{\rm s}(\vect r)=w \ast g^{12}(\vect r) \;.
\eea
As previously, the asterisk denotes a convolution.  
Thus the bridge function is
\bea
 \label{eq:bHHNC}
  b^{12}_{\rm HHNC}(\vect r) = \beta \mu^{\rm ex}(\rho_0\,g_{\rm s}(\vect r))- 
                  \beta \mu^{\rm ex}(\rho_0) +
                     \rho_0\, \tilde c^{(2)}(0)\,w 
           \ast h^{12}(\vect r)  \;.
\eea
The wall--fluid integral equation 
based on this bridge function,
\bea
 \label{eq:iehhnc}
 \log g^{12} +\beta u^{12} &=& \rho_0\,h^{12}\ast c^{(2)} - b^{12}_{\rm HHNC} 
\eea
supports a drying solution at the ``true'' coexisting 
densities. To see this, let $\rho_0=\rho_{\rm l}$, then within the thick drying film
$g^{12} = \rho_{\rm g}/\rho_{\rm l}$ and $\beta \mu^{\rm ex}
(\rho_0\,g_{\rm s}(\vect r))- \beta \mu^{\rm ex}(\rho_0)= 
\log( \rho_{\rm l}/\rho_{\rm g})$, consistent with equation (\ref{eq:iehhnc}). 
The drawback of this reasoning 
is that one needs explicitly the function $\mu^{\rm ex}(\rho)$ (which one would like
to actually obtain from the fluid integral equation) and the introduction of the
weight function is purely phenomenological. Furthermore, if one applies this
form of the bridge function to the integral equation of the fluid itself,
the results are clearly inferior to the present method or to reference HNC.
Thus there appears to be no route in HHNC to obtain the excess chemical potential
$\mu^{\rm ex}(\rho)$ self--consistently.

On the other hand, the formally exact bridge function  is given by
\bea
  b^{12} &=& \left.\beta \frac{\delta {\cal F}^{\rm B}}
   {\delta \rho(\vect r)}\right|_{\rho(\vect r)=\rho_{0}\; g^{12}(z)}
    \nonumber  \\
         &=& \beta\mu^{\rm ex}[\rho_0 g^{12}(z)]-
              \beta\mu^{\rm ex}(\rho_0)
         + \rho_0 h^{12}\ast c^{(2)}(z) \;,
\eea
which follows from equations (\ref{eq:fsplit}) and (\ref{eq:fhnc}).
Its approximation in the reference functional formalism is given by
\bea
 \label{eq:bFref}
  b^{12}       &\approx& \beta\mu^{\rm ex,ref}[\rho_0 g^{12}(z);d_1]-
              \beta\mu^{\rm ex,ref}(\rho_0;d_1)
         + \rho_0 h^{12}\ast c^{(2),\rm ref}(z;d_1) 
\eea  
Here the excess chemical potentials
$\mu^{\rm ex}[\cdot]$ and $\mu^{\rm ex,ref}[\cdot]$ are given  by
\bea
  \mu^{\rm ex[,ref]}[\rho(\vect r)] = \frac{\delta {\cal F}^{\rm ex[,ref]}}
   {\delta \rho(\vect r)}
\eea 
In the case of the Rosenfeld functional, $\mu^{\rm ex,ref}[\cdot]$ 
involves double convolutions 
with a certain set of geometric weight functions \cite{Ros89}. 
Comparing equations (\ref{eq:bHHNC})
and (\ref{eq:bFref}) we note that instead of employing  the exact fluid 
chemical potential $\mu^{\rm ex}$ and direct correlation function $c^{(2)}$, 
the respective reference system quantities appear and that the phenomenological
weight function $w$ is replaced by more complicated expressions related to the
nature of the reference functional. The comparison illustrates that  
compared to HHNC, the
bridge function within the reference functional method takes account of the
short--range correlation much better because of the use of 
the very accurate Rosenfeld functional. 
On the other hand, through the introduction of the reference system,
the hydrostatic limit is not fulfilled by the 
reference functional method; this is reflected by the fact that the coexisting
densities derived from hard--wall drying (equations (\ref{eq:deltamu}) 
and (\ref{eq:deltaom})) are nor perfectly consistent with those from  
the Maxwell construction.

\subsection{Sum rules and the relation to mean--field DFT}

For the planar hard wall problem, two important sum rules exist:
\bea
  \label{eq:prule}
  \frac{\beta p}{\rho_0} &=& g^{12}(0) \;, \\
  \label{eq:gam_rule}
  -\frac{\partial \gamma_{\rm wl}}{\partial \mu} &=& \Gamma_{\rm wl}(\mu) \;.
\eea 
The first is the contact sum rule (see equation (\ref{eq:prule1}))  which links the pressure $p$ in the bulk to the fluid density
at contact with the wall. The second is the Gibbs adsorption equation and links
the surface tension to the adsorption density. As usual, the derivative is 
performed on an isotherm.

\begin{figure}
\caption{\label{fig:dc} (colour online)
  The difference between  the direct correlation functions of the fluid
  and of the reference system for various stable liquid states.
}
 \vspace*{12mm}
 \begin{center}
   \epsfxsize=15cm
   \epsfbox{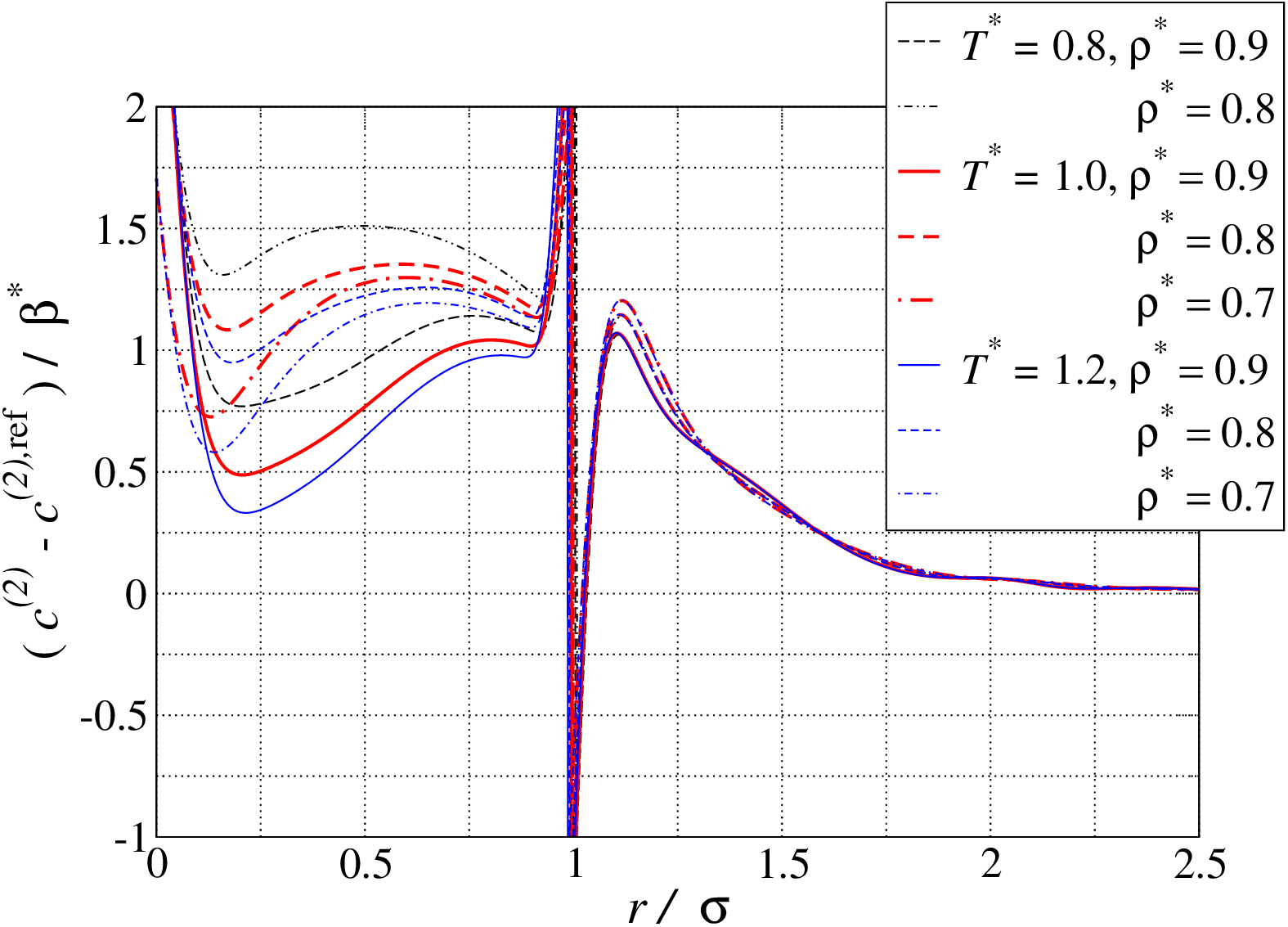}
 \end{center}
\end{figure}

The Gibbs adsorption rule remains valid for local density functionals and functionals 
which are constructed
from weighted densities (WDA).
The contact rule remains only valid for free energy functionals of WDA type
\cite{Swo89}. This  has rendered such functionals a popular tool for
the investigation of inhomogeneous matter. However, one should add the proviso
that in equation (\ref{eq:prule}), the pressure $p$ is the one derived from the
free energy functional for bulk densities, i.e. from the compressibility equation 
of state. 
Simple fluids are almost exclusively treated
in density functional theory using a mean--field approximation for the 
attractive tails (see below)
for which the consistency between the virial and compressibility equation of state
is not very good.  Moreover, one might 
object that agreement with sum rule (\ref{eq:gam_rule}) does not guarantee
the accuracy of $\gamma_{\rm wl}$ and $\Gamma_{\rm wl}$ themselves,
as these quantities depend strongly  on the quality of the free energy functional.

\begin{figure}
\caption{\label{fig:sumrule1} (colour online)
 Testing the consistency of the contact density sum rule, equation (\ref{eq:prule}). 
 We have plotted the ratio $p^\ast/(\rho_0^\ast\,g^{12}(0))$ for various temperatures
 $T^\ast =0.8\dots 1.4$. The sum rule value of this ratio, $T^\ast$, is shown
 by a horizontal straight line. For the three lowest temperatures,
 results are plotted for the stable liquid phase only.
}
 \vspace*{13mm}
 \begin{center}
   \epsfxsize=12cm
   \epsfbox{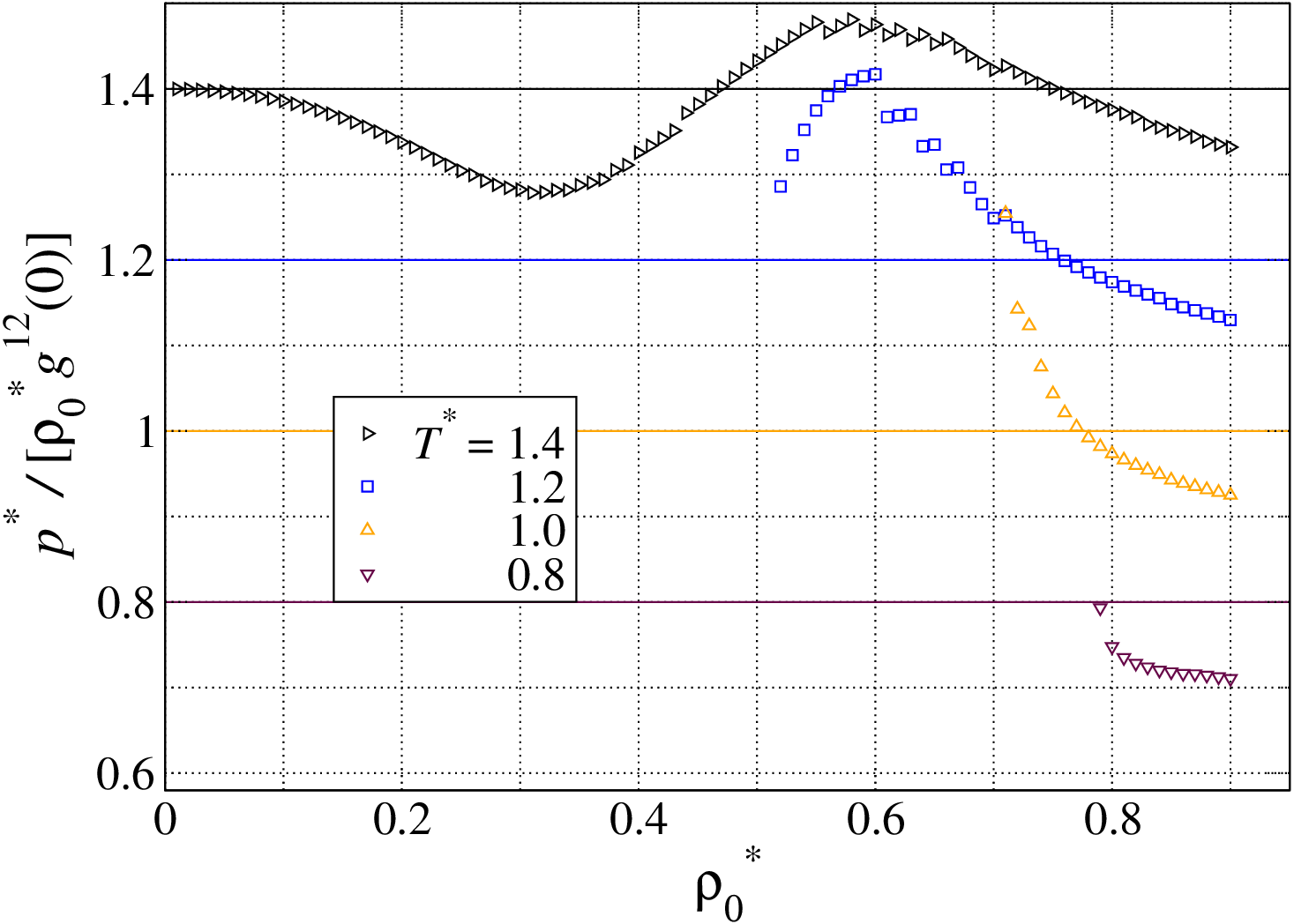}
 \end{center}
\end{figure}

\begin{figure}
\caption{\label{fig:sumrule2} (colour online)
 Testing the consistency of the Gibbs adsorption rule, equation (\ref{eq:gam_rule}). 
 We show a comparison of the adsorption density $\Gamma_{\rm wl}=\rho_0\int dz h^{12}$
 (symbols) with the same quantity obtained from sum rule (\ref{eq:gam_rule})
 as a function of the bulk density
 for various temperatures
 $T^\ast =0.8\dots 1.4$. To perform the $\mu$ derivative of the surface tension
 $\gamma_{\rm wl}$, both $\mu$ and $\gamma_{\rm wl}$ have been fitted to polynomials
 in $\rho_0$. The fit is of limited accuracy only very close to coexistence. 
 The vertical dashed lines for the three lowest temperatures mark the coexisting
 liquid densities.
}
 \vspace*{10mm}
 \begin{center}
   \epsfxsize=7.3cm
   \epsfbox{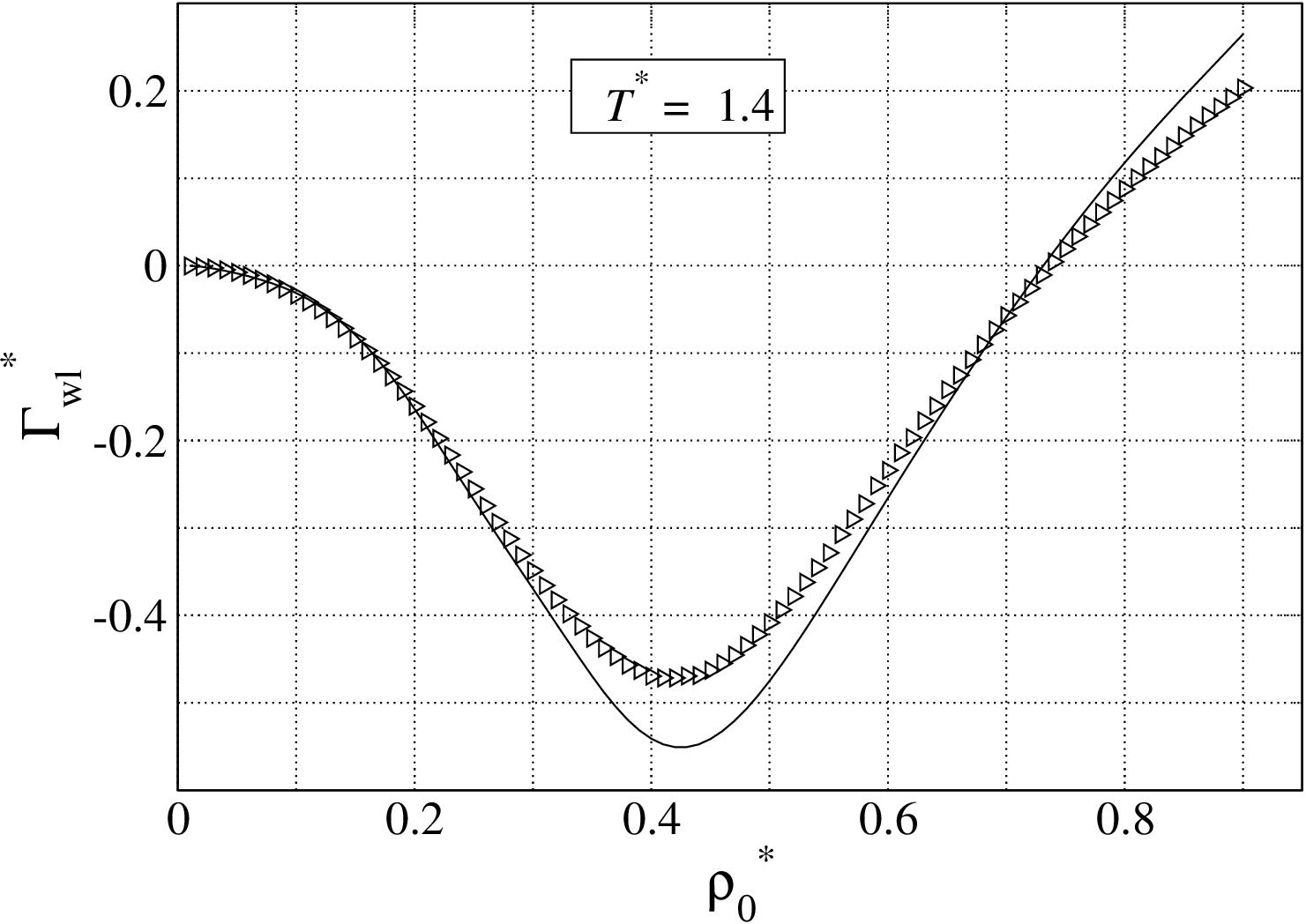} \hspace{0.6cm}
   \epsfxsize=7.3cm
   \epsfbox{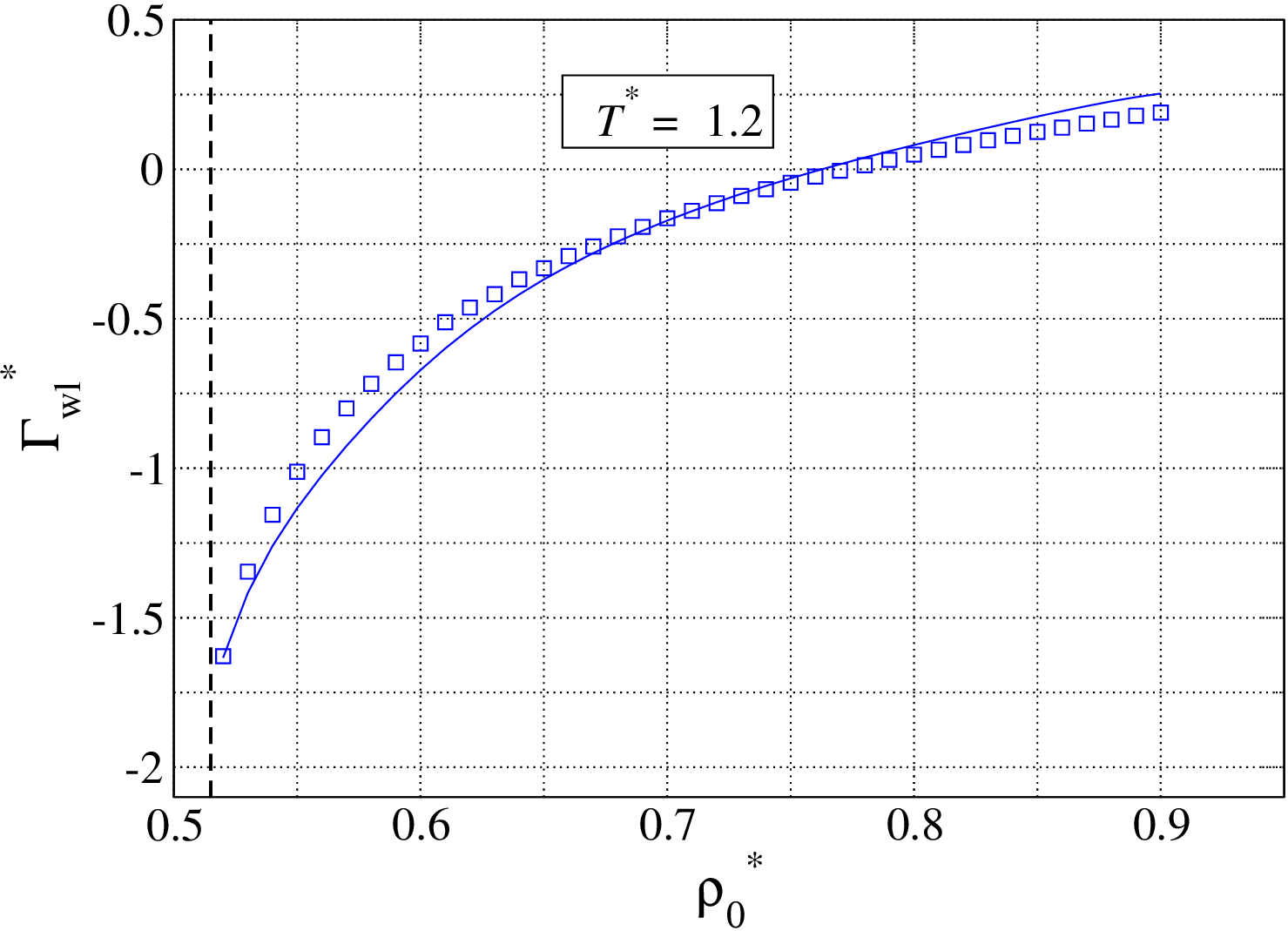} \\[2cm]
   \epsfxsize=7.3cm
   \epsfbox{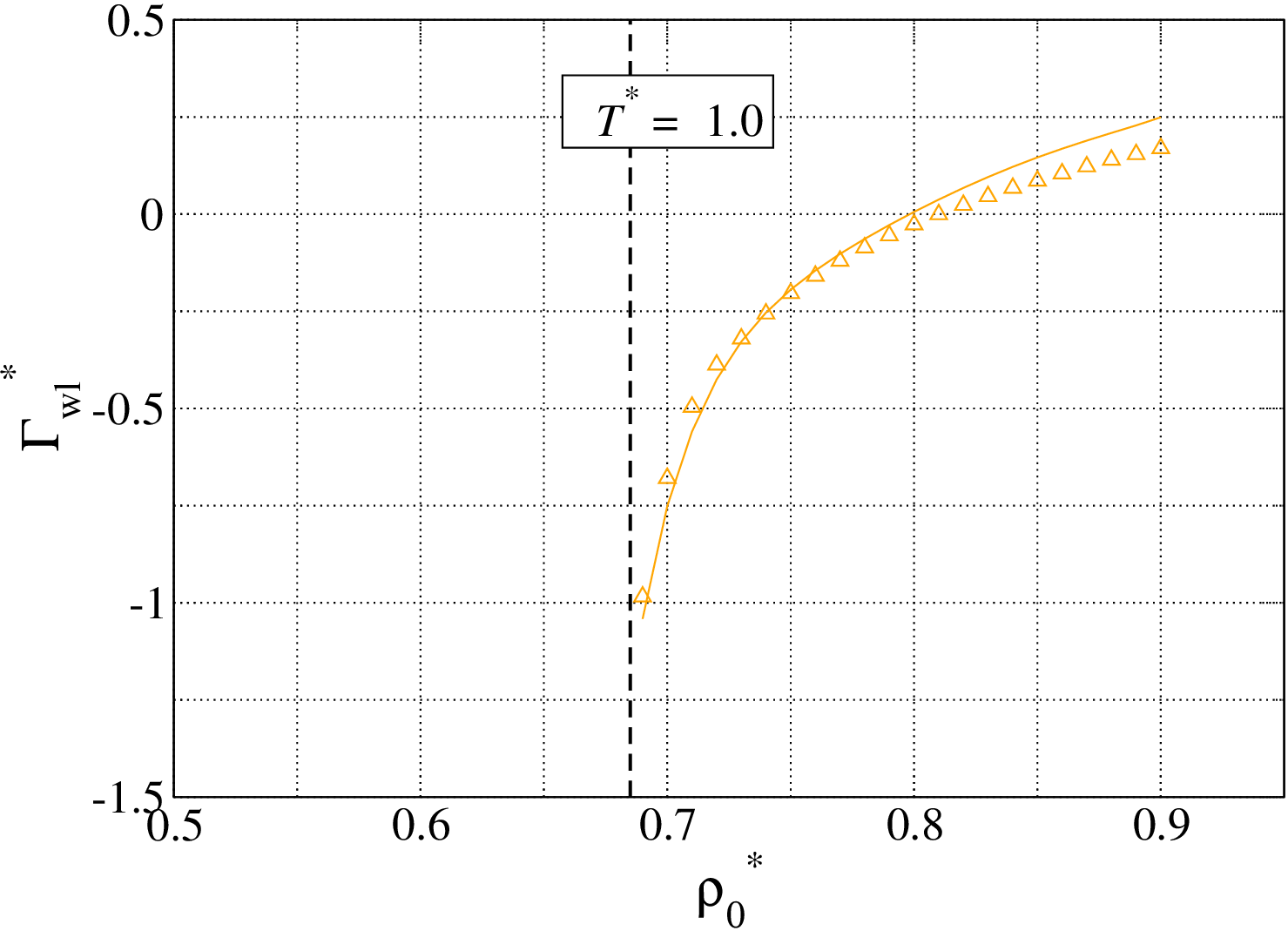} \hspace{0.6cm}
   \epsfxsize=7.3cm
   \epsfbox{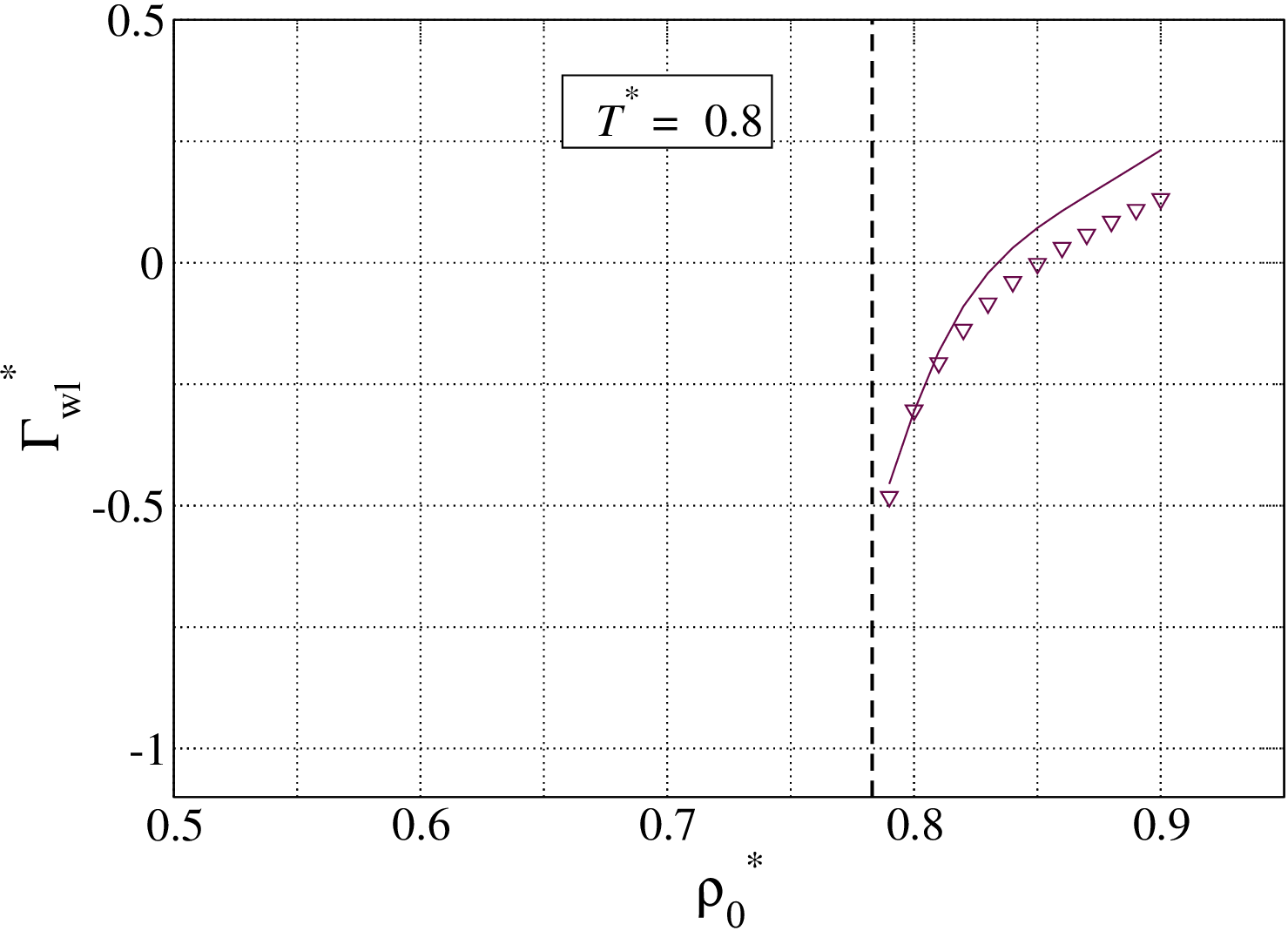} 
 \end{center}
\end{figure}

For hard spheres, these problems are solved to a high degree of accuracy
by presently available functionals \cite{Rot02,Bry03}. The simplest mean--field
treatment of fluids in which the pair potential $u^{11}(r)$ has an attractive tail 
is given by
\bea
  \label{eq:mf}
  {\cal F}^{\rm ex}[\rho(r)] &=& {\cal F}^{\rm ex,ref}[\rho(r);d_1={\rm const.}]
   + \frac{1}{2} \int d\vect r\; \rho(\vect r)\; w_{\rm att}\ast \rho(\vect r) \; ,
\eea 
where we consider a one--component system.
Here, the weight function $w_{\rm att}$ is given by  the attractive tail
of the interparticle potential (for which various recipes exist in the literature). 
If we compare the second term in equation (\ref{eq:mf})
with the excess free energy beyond the reference system in our approach 
(see equation (\ref{eq:mf_gen})), then the reference functional approach predicts
roughly mean--field behaviour for the {\em density deviations} from the bulk.
Thus we may conclude, if the prefactor of the quadratic term in equation 
(\ref{eq:mf_gen}), 
$(c^{(2)}-c^{(2),{\rm ref}})/\beta$, does not vary appreciably with 
the bulk density $\rho_0$ then a description in terms of the functional
given in equation (\ref{eq:mf}) will capture the essential physics. 
Checking the numerical results (see Figure \ref{fig:dc}), we find that 
for $r>\sigma$ indeed a single function describes this prefactor 
for all stable liquid state points except for the ones in the critical region. 
Inside the harshly repulsive core ($r<\sigma$) we notice some variation
which is, however, small when compared to the magnitude of the direct pair correlation
function of the reference system. 
Thus
the ``mean--field'' expansion around the bulk density
as given in equation (\ref{eq:mf_gen}) is expected to entail similar physics
as that in the mean--field functional (\ref{eq:mf}) 
 but the former has the effect of 
giving a more accurate equation of state and very reliable  fluid correlation
functions. The mean field approach (\ref{eq:mf_gen}) is, in turn, computationally less demanding
and has the advantage that the sum rules hold analytically (with respect to the compressibility
equation of state). 
  
It is important to note that the very concept of expanding around a fixed bulk density 
implies that
the sum rule proofs which work for weighted density functionals do not hold
for the Taylor expanded functional \cite{Swe00}. Therefore we have tested both sum rules
numerically for various temperatures.
In Figure \ref{fig:sumrule1} we have plotted the ratio of the virial
pressure to the density at contact, $p^\ast/[\rho_0 g^{12}(0)]$ (which should
give $T^\ast$), for temperatures ranging from $T^\ast=0.8 \dots 1.4$.
The deviations are up to 20\%. 
However, despite these deviations these results are 
encouraging since here the {\em virial} ($=$ quasi--exact) pressure is used.
In Figure \ref{fig:sumrule2} we compare the adsorption densities
calculated by equation (\ref{eq:Gamdef}) with the adsorption density obtained
from the surface tension via the sum rule (\ref{eq:gam_rule}).
Again we find very reasonable agreement. 

\section{Summary and conclusions}
\label{sec:summary}

In this article, we have analyzed the pair correlation functions of an inhomogeneous
fluid mixture by means of a functional Taylor expansion of the free energy
around an inhomogeneous density profile.
We have derived a general integral equation closure
for inhomogeneous systems which is based on the introduction of a reference
functional for the excess free energy beyond second order in the Taylor
expansion. A benefit of this approach is that explicit expressions
(see equation (\ref{eq:c1_ins})) for
the insertion free energies of particles into the inhomogeneous
equilibrium distribution are obtained, which reduce to explicit expressions for the
excess chemical potential if the expansion is performed around a 
homogeneous (bulk) state.

As an application, we have considered the equation of state for a bulk cut--off
and shifted Lennard--Jones fluid and  the structure and adsorption
of this fluid at a hard wall. Treating this
problem within the mixture theory and expanding around the bulk state, 
we find excellent agreement between  the virial
and internal energy routes to the  equation of state for the fluid  and with  
simulations. 
Thermodynamic consistency is good for stable state points outside the critical region.
By virtue of the underlying hard sphere reference functional, complete drying
is predicted for the hard wall--liquid interface and the corresponding coexisting 
densities for the liquid branch are in excellent agreement with those from 
the Maxwell construction. However, there are  deviations for the coexisting 
densities on the
gas branch. Nevertheless, this reasonable consistency between the 
Maxwell construction and hard wall drying is a novel result within integral 
equation approaches.  

Two sum rules related to the hard wall problem, the contact density rule,
equation (\ref{eq:prule}), and the Gibbs adsorption equation (\ref{eq:gam_rule}), 
have been investigated over a wide range of densities
and temperatures. Although these sum rules are not satisfied exactly, the deviations
are reasonably small, especially near coexistence. 

The numerical results for both the fluid--fluid and the wall-fluid correlations suggest 
a generalized mean field behavior outside the critical region. 
In the reference functional approach, the 
difference between the fluid free energy and the free energy of the reference system
is expanded up to second order in the density difference (from the bulk density),
see equation (\ref{eq:mf_gen}). The coefficient of the second--order term, the difference 
between the pair direct correlation functions of the fluid and the reference
system, is determined self--consistently and turns out to be quite insensitive
to variations in temperature and bulk density for liquid states, see Figure
\ref{fig:dc}.  
Therefore we argue
that for physical problems related to wetting and drying
a description with 
density functionals composed of an accurate reference part and  a mean field 
attractive term should be sufficient to capture all the essential features 
and will be much less demanding in computational
effort. 

Related to the efficacy of generalized mean--field treatments of simple fluids,
we note that a quite different mean--field treatment  of the density deviations from the bulk 
was already developed in reference \cite{Wee98} in which the
behaviour of  a LJ fluid at a hard wall was studied as well. In contrast to the
present study, the properties of the hard sphere reference system were fixed
by considering the first YBG equation, for details see reference \cite{Wee98}.
Good agreement between simulated and calculated density profiles was obtained
for selected supercritical state points.

We anticipate interesting perspectives for further computations in the problem of
small colloidal particles at interfaces. The free energy profile for particles dragged
through the interface or their mutual interactions at the interface are 
only beginning to be understood. These observables can be analyzed within
the reference functional approach for inhomogeneous distributions;
the challenging numerical implementation remains a task for future work.

\ack
 The author is grateful to the Alexander von Humboldt--Foundation for a 
research grant supporting a stay in Bristol during which a good part of this work
was completed.
He thanks the Physics Department in Bristol and especially Bob Evans and Andy Archer 
for their hospitality and for enjoyable discussions. Bob Evans has followed 
the development of the work presented here with great interest and has provided 
many helpful suggestions  and he is also
thanked for a thorough reading of the manuscript. Furthermore, the author thanks
Roland Roth for beneficial discussions.

\appendix
\setcounter{section}{0}
\section*{Appendix}

\section{Surface tension at the hard wall}
\label{app:gamma}

The hard wall--liquid surface tension is defined by the expression
\bea
  \gamma_{\rm wl} = \int_0^\infty dz (\omega(z)+p)\;,
\eea
where $\omega(z)$ is the grand potential volume density and $p=p(\rho_0)$
is the bulk pressure of the liquid with density $\rho_0$. In terms
of the grand potential density functional, this equation is equivalent to
\bea
 \label{eq:gwl_def}
  \gamma_{\rm wl} =\left(\left.\Omega[{ \rho}(\vect r)]
  \right|_{ \rho(\vect r) = \rho_0\,g^{12}(z)} - 
    \left.\Omega[{\rho(\vect r)}]\right|_{\rho(\vect r) = \rho_0\,\theta(z)}
   \right)/A\;,
\eea
where $A$ denotes the area of the wall. We use the definition
\bea
  \Omega[{ \rho}(\vect r)] = {\cal F}[\rho(\vect r)] - 
  \int d\vect r (\mu(\rho_0)-\xi\,u^{12}(\vect r))\,\rho(\vect r)\;.
\eea
with $\xi=1$ for the first term in equation (\ref{eq:gwl_def}) and
$\xi=0$ for the second term. According to the discussion
below equation (\ref{eq:h12}) concerning the hard wall limit,
the functional of the free energy
is the functional for the one--component LJ fluid,
\bea
   {\cal F}[\rho(\vect r] = {\cal F}^{\rm id}[\rho(\vect r)]
  + {\cal F}^{\rm HNC}[\rho(\vect r)] +  {\cal F}^{\rm B,ref}[\rho(\vect r);d_1]\;,
\eea
with $d_1=d_1(\rho_0)$ denoting the reference system hard sphere diameter.
Straightforward evaluation gives
\bea
 \label{eq:a1}
\fl \beta \gamma_{\rm wl} &=& \int_0^\infty dz\,\rho_0\left( g^{12}\ln g^{12}
   - h^{12}(1+\beta \mu^{\rm ex,ref}(\rho_0)) -\frac{\rho_0}{2} 
   \left[(c_z^{(2)}-c_z^{(2),{\rm ref}})\ast  h^{12}\right]\, h^{12} \right) + 
  \nonumber \\ 
\fl  & &   \beta{\cal F}^{\rm ex,ref}[\rho_0\,g^{12}] - 
  \beta\int_0^\infty dz\, f^{\rm ex,ref} + I_{\rm wall} \;, \\
\fl  I_{\rm wall} &=&  \int_{-\infty}^0 dz\,\left( \beta[ f^{\rm ex}(\rho_0)-f^{\rm ex,ref}(\rho_0)]+
   \phantom{ \frac{1}{2}} \right.\\
\fl & & \qquad \qquad \left.  
   \rho_0 h^{12} \beta (\mu^{\rm ex}(\rho_0)- \mu^{\rm ex,ref}(\rho_0)) 
   -\frac{\rho_0^2}{2}
   \left[(c_z^{(2)}-c_z^{(2),{\rm ref}})\ast  h^{12}\right]\, h^{12} \right)\;. \nonumber
\eea
Here, $f^{\rm ex}(\rho_0)$ and $f^{\rm ex,ref}(\rho_0)$ are the bulk 
free energy volume densities 
of the fluid and the hard--sphere reference system, respectively. 
We note that the non--local expansion of the free energy in terms of 
$\Delta\rho=\rho_0\,h^{12}$ ($=-\rho_0$ for $z<0$) introduces 
contributions inside the wall, contained in  $I_{\rm wall}$.
Furthermore,
\bea
  c_z^{(2)[,{\rm ref}]} (z) = 2\pi\int_{|z|}^\infty r dr\, c^{(2)[,{\rm ref}]} (r)\;,
\eea
and we have introduced the one--dimensional convolution
\bea
 c_z^{(2)[,{\rm ref}]} \ast h^{12}\,(z) = \int_{-\infty}^\infty dz' 
   c_z^{(2)[,{\rm ref}]}(z-z') h^{12}(z') \; .
\eea
The unknown free energy density of the liquid $f^{\rm ex}(\rho_0)$ 
appearing in the integral $I_{\rm wall}$ is fixed by the requirement
that beyond the wall ($z<0$) the density is zero and therefore also $f^{\rm ex}(0)=0$.
On the other hand, $f^{\rm ex}(0)$ is determined by the functional Taylor expansion 
of the fluid excess
free energy around $\rho_0$, evaluated
at zero density. Therefore we find:
\bea
\fl  f^{\rm ex}(\rho_0) =  f^{\rm ex,ref}(\rho_0) + \rho_0\,(\mu^{\rm ex}(\rho_0)- 
  \mu^{\rm ex,ref}(\rho_0))  + \frac{\rho_0^2}{2}
  \left[ \tilde c^{(2)}(0)-\tilde c^{(2),{\rm ref}}(0)\right]\;.
\eea
This expression is inserted into $I_{\rm wall}$ and we obtain the expression:
\bea
\fl  I_{\rm wall} &=&  \frac{\rho_0^2}{2}\int_{-\infty}^0 dz\,\left(
  \left[ \tilde c^{(2)}(0)-\tilde c^{(2),{\rm ref}}(0)\right] -
    \left[(c_z^{(2)}-c_z^{(2),{\rm ref}})\ast  h^{12}\right]\, h^{12} \right) \; . 
\eea
This completes the prescription for calculating $\gamma_{\rm wl}$.

\end{document}